\documentclass[12pt]{article}
\pdfoutput=1

\usepackage{amsmath}
\usepackage{amssymb}
\usepackage{graphicx}
\usepackage{cite}
\usepackage{multirow}
\usepackage{longtable}
\usepackage{lscape}

\allowdisplaybreaks[4] 

\newcommand{\capdef}{}
\newcommand{\mycaption}[2][\capdef]{\renewcommand{\capdef}{#2}%
        \caption[#1]{{\footnotesize #2}}}
\makeatletter
\renewcommand{\fnum@table}{\textbf{\tablename~\thetable}}
\renewcommand{\fnum@figure}{\textbf{\figurename~\thefigure}}
\makeatother

\newcommand{\ie}{{\it i.e.}}

\newcommand{\cf}{{\it cf.}}

\newcommand{\eq}{Eq.}

\newcommand{\fig}{Fig.}

\newcommand{\Ref}{Ref.}
\newcommand{\Refs}{Refs.}
\newcommand{\Sec}{Sec.}

\newcommand{\App}{Appendix}

\newcommand{\Tab}{Table}

\newcommand{\equ}[1]{\eq~(\ref{equ:#1})}
\newcommand{\figu}[1]{\fig~\ref{fig:#1}}

\newcommand{\bi}{\begin{itemize}}
\newcommand{\ei}{\end{itemize}}

\newcommand{\be}{\begin{equation}}
\newcommand{\ee}{\end{equation}}
\newcommand{\bea}{\begin{eqnarray}}
\newcommand{\eea}{\end{eqnarray}}

\newcommand{\ldm}{\Delta m_{31}^2}
\newcommand{\sdm}{\Delta m_{21}^2}

\newcommand{\deltacp}{\delta_{\mathrm{CP}}}
\newcommand{\stheta}{\sin^2 2 \theta_{13}}

\newcommand{\br}{$^8$B~}
\newcommand{\li}{$^8$Li~}
\newcommand{\he}{$^6$He~}
\newcommand{\neon}{$^{18}$Ne~}

\newcommand{\DC}{{\sc DeepCore}}
\newcommand{\dc}{{\sc PINGU}}

\newcommand{\BBref}{BB$_{\text{ref}}$}
\newcommand{\BBrefs}{BB$_{\text{ref}}^*$}
\newcommand{\NFref}{NF$_{\text{ref}}$}
\newcommand{\NFrefs}{NF$_{\text{ref}}^*$}
\newcommand{\SBrefs}{SB$_{\text{ref}}^*$}
\newcommand{\BBbm}{BB$_{\text{bm}}$}
\newcommand{\BBbms}{BB$_{\text{bm}}^*$}
\newcommand{\NFbm}{NF$_{\text{bm}}$}
\newcommand{\NFbms}{NF$_{\text{bm}}^*$}
\newcommand{\SBbms}{SB$_{\text{bm}}^*$}

\begin{document}
\begin{titlepage}

\renewcommand{\thefootnote}{\alph{footnote}}

\vspace*{-3.cm}
\begin{flushright}
EURONU-WP6-11-42  \\
IDS-NF-031 \\
\end{flushright}


\renewcommand{\thefootnote}{\fnsymbol{footnote}}
\setcounter{footnote}{-1}

{\begin{center}
{\large\bf
Requirements for a New Detector at the South Pole \\
Receiving an Accelerator Neutrino Beam\\[0.3cm]
} \end{center}}
\renewcommand{\thefootnote}{\alph{footnote}}

\vspace*{.3cm}
{\begin{center}
{\large \today}
\end{center}
}

{\begin{center} {\large{\sc
 		 Jian~Tang\footnote[1]{\makebox[1.cm]{Email:}
                 jtang@physik.uni-wuerzburg.de} and
                 Walter~Winter\footnote[2]{\makebox[1.cm]{Email:}
                 winter@physik.uni-wuerzburg.de}}
		 }
\end{center}
}
\vspace*{0cm}
{\it
\begin{center}

\footnotemark[1]${}^,$\footnotemark[2]
       Institut f{\"u}r Theoretische Physik und Astrophysik, 
       Universit{\"a}t W{\"u}rzburg, \\
       D-97074 W{\"u}rzburg, Germany

\end{center}}

\vspace*{0.5cm}

{\Large \bf
\begin{center} Abstract \end{center}  }

There are recent considerations to increase the photomultiplier density in the IceCube detector array 
beyond that of DeepCore, which will lead to a lower detector energy threshold and a relatively huge fiducial mass for the neutrino detection. This initiative is known as ``Precision IceCube Next Generation Upgrade'' (\dc). We discuss the possibility to send a neutrino beam from one of the major accelerator laboratories in the Northern hemisphere to such a detector. Such an experiment would be unique in the sense that it would be the only neutrino beam where the baseline crosses the Earth's outer core.
We study the detector requirements for  a beta beam, a neutrino factory beam, and a superbeam, where we consider the cases of both small $\theta_{13}$ and large $\theta_{13}$, as suggested by the recent T2K and Double Chooz results. 
We illustrate that a flavor-clean beta beam best suits the requirements of such a detector, in particular, that PINGU may replace a magic baseline detector for small values of $\theta_{13}$ -- even in the absence of any energy resolution capability. For large $\theta_{13}$, however, a single-baseline beta beam experiment cannot compete if it is constrained by the CERN-SPS. For a neutrino factory, because of the missing charge identification possibility in the detector, a very good energy resolution is required. If this can be achieved, especially a low energy neutrino factory, which does not suffer from the tau contamination, may be an interesting option for large $\theta_{13}$.
For the superbeam, where we consider the LBNE beam as a reference, electron neutrino flavor identification and statistics are two of the primary limitations.
Finally, we demonstrate that in principle the neutrino factory and superbeam options may measure the density of the Earth's core at a sub percent level for $\stheta \gtrsim 0.01$.

\vspace*{.5cm}

\end{titlepage}

\newpage

\renewcommand{\thefootnote}{\arabic{footnote}}
\setcounter{footnote}{0}

\section{Introduction}

Neutrino oscillation experiments have provided compelling evidence
that the weakly interacting neutrinos are massive
particles~\cite{GonzalezGarcia:2007ib}. Given three generations of massive neutrinos, there must be two characteristic mass squared splittings ($\Delta m_{31}^2\,,\Delta m_{21}^2$) and three mixing angles ($\theta_{12}\,,\theta_{13}\,,\theta_{23}$) as well as a Dirac-type CP violation phase ($\delta_{\textrm{CP}}$) affecting neutrino oscillations.  Disappearance of muon neutrinos, which is mainly driven by $|\Delta m_{31}^2|$ and $\theta_{23}$, has been observed in atmospheric neutrino oscillation experiments, such as Super-Kamiokande~\cite{Fukuda:1998mi}, and in the MINOS long baseline experiment~\cite{Adamson:2008zt}. Disappearance of electron neutrinos has been observed from solar neutrino oscillation experiments relating to $\theta_{12}$~\cite{Ahmad:2002jz}, whereas  $\Delta m_{21}^2$ has been strongly constrained by the KamLAND long baseline reactor neutrino experiment~\cite{Araki:2004mb}. The CHOOZ short-baseline reactor neutrino oscillation experiment~\cite{Apollonio:2002gd} has provided a limit $\stheta \lesssim 0.1$. Recently there has been a $2.5\sigma$ hint for non-zero $\theta_{13}$ from the T2K experiment~\cite{Abe:2011sj} and the Double Chooz experiment~\cite{Abe:2011fz}, see \Refs~\cite{Fogli:2011qn,Schwetz:2011zk} for global fits. 
Since it is too early to claim a $\theta_{13}$ discovery, we leave $\theta_{13}$ as unknown parameter in this study. We qualitatively distinguish the ``large $\theta_{13}$'' ($\theta_{13}$ discovered by the next generation of experiments) and ``small $\theta_{13}$'' ($\theta_{13}$ constrained by the next generation of experiments) cases. These cases are roughly separated by $\stheta=0.01$~\cite{Huber:2004ug,Huber:2009cw}.
The most important questions are then: is  $\Delta m_{31}^2>0$ (normal ordering) or $\Delta m_{31}^2<0$ (inverted ordering), what is the value of $\theta_{13}$, and is there leptonic CP violation (CPV). 
Other open questions concern the possible existence of light sterile neutrinos, as indicated by recent MiniBOONE antineutrino results~\cite{AguilarArevalo:2010wv} confirming the LSND anomaly~\cite{Aguilar:2001ty}, and a (possibly energy- or environment-) dependent superluminal motion of neutrinos~\cite{Adam:2011zb}.
 Many of these questions can most likely not be addressed with existing experiments~\cite{Huber:2009cw}, including the reactor experiments searching for $\theta_{13}$. A next generation of experiments, using accelerator-based technology in the Northern hemisphere, has been studied in international efforts, including the International Neutrino Factory and Superbeam Scoping Study~\cite{Bandyopadhyay:2007kx,Abe:2007bi,Berg:2008xx}, the ongoing International Design Study for the Neutrino Factory (IDS-NF)~\cite{ids}, and the Euronu design study ``A High Intensity Neutrino Oscillation Facility in Europe''~\cite{euronu}. 

In the Southern hemisphere, the IceCube Neutrino Observatory~\cite{Ahrens:2003ix} has recently completed deployment at the South Pole, including an array with denser  string configuration called \DC\ with a lower threshold $\mathcal{O}(10)$~GeV. Given that this technology is very well established and the cost for future potential upgrades is relatively well predictable, there is recent activity for the extension of IceCube in the low energy direction, called \dc\ (Precision IceCube Next Generation Upgrade). The primary physics applications for this experiment are those similar to that of other megaton-size water Cherenkov (WC) detectors, such as proton decay, detection of supernova neutrinos, and detection of atmospheric neutrinos. However, such a detector could also conceivably receive a neutrino beam, as discussed for some superbeam-upgrade and beta beam proposals in the Northern hemisphere (for example, CERN-Frejus or J-PARC to Hyper-Kamiokande). For the first step of such an upgrade, 20 additional strings are discussed leading to a fiducial mass of 1~Mt at a few GeV, increasing to about 10~Mt at 10~GeV. In a future phase, for which the detector parameters are not yet specified, there may be even a substantially larger fiducial mass. Although the baseline from any of the major accelerator laboratories in the Northern hemisphere is known to be suboptimal for neutrino oscillation studies, in particular for CP violation, we study in this work if the large increase of statistics by this relatively huge detector mass may compensate for the loss of statistics from the long baseline. More specifically, consider a beta beam, neutrino factory beam, and a superbeam, and establish the detector requirements for each of these beam classes for small and large $\theta_{13}$. \Sec~\ref{sec:detector} discusses our detector parameterization. These requirements may serve as guidance for the optimization of the detector, if it were intended to receive a neutrino beam.
Our primary focus is the optimization with respect to the most often used performance indicators, including the $\theta_{13}$, mass hierarchy (MH), and CP violation (CPV) discovery reaches as a function of (true) $\theta_{13}$ and $\deltacp$. Note that the neutrino oscillation physics using such a long core-crossing baseline is quite unique because of the ``parametric enhancement''~\cite{Akhmedov:1998ui,Chizhov:1998ug} of the oscillation probability; for a detailed discussion, see \Sec~\ref{sec:oscillation}.
Conversely, the density of the Earth's core can be probed with such a neutrino beam~\cite{Winter:2005we}, which we study in \Sec~\ref{sec:earth}.
Earlier works studying a neutrino beam to the South Pole include \Refs~\cite{Dick:2000fn,Fargion:2010vb}.

As far as the different beam classes are concerned, we distinguish setups for small $\theta_{13}$, where only an upper bound on $\theta_{13}$ can be expected from the next generation of superbeam and reactor experiments, and large $\theta_{13}$, where $\theta_{13}$ will be discovered by the next generation of experiments. The setups for small $\theta_{13}$ should be able to access as small as possible $\theta_{13}$, whereas the setups for large $\theta_{13}$  should have an optimal $\deltacp$ reach. For a neutrino factory, the small and large $\theta_{13}$ setups are specified in the IDS-NF Interim Design Report~\cite{IDSNF:2011aa}. For small values of $\theta_{13}$, a two-baseline setup with $L_1 \simeq 3000$--$5000 \, \mathrm{km}$ (for CP violation) and the so-called ``magic baseline''~\cite{Huber:2003ak} with $L_2 \simeq 7500 \, \mathrm{km}$ (for degeneracy resolution and the mass hierarchy measurement) is proposed, with a muon energy $E_\mu = 25 \, \mathrm{GeV}$. For large values of $\theta_{13}$, a single-baseline setup with a smaller muon energy and shorter baselines is preferable, where the exact parameters depend on the detector properties. For a magnetized iron detector, $E_\mu \simeq 10 \, \mathrm{GeV}$ and $L \simeq 2200 \, \mathrm{km}$ has been obtained in \Ref~\cite{Agarwalla:2010hk}.
For a beta beam, qualitatively similar setups have been established for small~\cite{Choubey:2009ks} and large~\cite{Winter:2008cn} values of $\theta_{13}$, using two and one baselines, respectively. For a superbeam, typically only single-baseline options are considered in the literature since the intrinsic contamination of the beam with electron neutrinos limits the $\theta_{13}$ reach. An example of this is the Long-Baseline Neutrino Experiment (LBNE) wide band beam experiment from the Fermi National Accelerator Laboratory (FNAL, Batavia, Illinois) to Homestake (North Dakota) in the US~\cite{LBNE}. In this study, we consider modifications of the above beam setups. For small values of $\theta_{13}$, we test if the magic baseline detector for a beta beam or neutrino factory could be replaced by \dc ; see \Sec~\ref{sec:two}. For large values of $\theta_{13}$, we consider single-baseline options using a beta beam, neutrino factory beam, or superbeam; see \Sec~\ref{sec:one}. Details on the beams, detector requirements, and reference setups can be found in \Sec~\ref{sec:reference}. 

\section{Oscillation physics with a core-crossing baseline}
\label{sec:oscillation}

The relevant information on $\stheta$, $\deltacp$, and the mass hierarchy at long-baseline neutrino oscillation experiments is typically inferred from the appearance probability $P_{e \mu} \equiv P_{\nu_e \rightarrow \nu_\mu}$ (or $P_{\mu e}$). For a constant density profile, it is typically
appropriate to expand this probability to second order in $\sin 2 \theta_{13}$ and the hierarchy parameter $\alpha \equiv \sdm/\ldm \simeq 0.03$ as~\cite{Cervera:2000kp,Freund:2001pn,Akhmedov:2004ny}\footnote{Although this expansion is not particularly accurate for very long baselines~\cite{Akhmedov:2004ny}, especially for baselines crossing the Earth's core, it is sufficient to illustrate our main qualitative points.}
\begin{eqnarray}
 P_{e\mu} &\simeq& 
 \sin^22\theta_{13} \sin^2\theta_{23} 
\frac{\sin^2[(1-\hat{A})\Delta_{31}]}{(1-\hat{A})^2}\nonumber \\
&\pm& \alpha \sin2\theta_{13} \sin\deltacp \sin2\theta_{12} \sin2\theta_{23}  \sin(\Delta_{31}) \frac{\sin(\hat{A}\Delta_{31})}{\hat{A}}
\frac{\sin[(1-\hat{A})\Delta_{31}]}{(1-\hat{A})} \nonumber \\
&+& \alpha \sin2\theta_{13} \cos\deltacp \sin2\theta_{12} \sin2\theta_{23}  \cos(\Delta_{31}) \frac{\sin(\hat{A}\Delta_{31})}{\hat{A}}
\frac{\sin[(1-\hat{A})\Delta_{31}]}{(1-\hat{A})} \nonumber \\
&+& \alpha^2 \cos^2\theta_{23} \sin^22\theta_{12} 
\frac{\sin^2(\hat{A}\Delta_{31})}{{\hat{A}}^2} \, .
\label{equ:papp}
\end{eqnarray}
Here $\Delta_{31} \equiv \ldm L/(4E)$, $\hat A=2 \sqrt{2} G_F n_e E/\ldm$ with $n_e$ the electron density in Earth matter, and the sign of the second term refers to neutrinos (positive) or anti-neutrinos (negative).
Note that the sign of $\hat{A}$ also depends on considering
neutrinos or anti-neutrinos. The T-inverted probability,
$P_{\mu e}$, may be obtained from \equ{papp} by changing only the
sign of the second term. 

Especially if $ \sin 2 \theta_{13} \sim \alpha$, \ie, all the four terms in \equ{papp} contribute equally, the correlations and degeneracies affect the extraction of the individual parameters~\cite{Barger:2001yr}. A convenient method is to combine two baselines~\cite{IDSNF:2011aa,Choubey:2009ks}:
a shorter baseline sensitive to all four terms in \equ{papp} to measure CP violation (in the second and third terms), and a longer ``magic'' baseline~\cite{Huber:2003ak} $L \simeq 7 \, 500 \, \mathrm{km}$ for degeneracy resolution. At this second baseline $\sin ( \hat A \Delta_{31} ) = 0$, meaning that only the first term in \equ{papp} survives allowing for a clean $\stheta$ and mass hierarchy measurement. The condition $\sin ( \hat A \Delta_{31} ) = 0$ can be fulfilled independent of oscillation parameters and neutrino energy, and depends only on the matter density profile. However, if two baselines are combined, a wider baseline range produces a similar effect~\cite{Gandhi:2006gu,Kopp:2008ds}. The large fiducial mass of \dc\ may compensate for any disadvantage from a longer baseline, such as lower event rates and some re-occurring dependence on $\deltacp$ (which may prove useful for CPV studies). Therefore, we will study two-baseline setups where the second baseline is replaced by that to \dc . For large values of $\theta_{13}$, the first term in \equ{papp} dominates, and is the primary background for the CP violation measurement. Therefore, systematic uncertainties are important, but their detailed investigation goes beyond this study. Note, however, that broader beam spectra~\cite{Barger:2007jq}, or longer baselines that include more oscillation peaks~\cite{Coloma:2011pg} tend to be less sensitive to systematics than narrow-band off-axis beams operated at the first oscillation maximum.

Another peculiarity obtained from \equ{papp} is the effect of the MSW resonance $\hat A \rightarrow 1$. In this case the first term will be minimally affected by the $1/L^2$ decrease of the flux for a wide baseline range~\cite{Freund:1999gy}, meaning that large event rates can be expected even for very long baselines. This condition, however, is energy-dependent, and is satisfied for the MSW resonance energy
\begin{equation}
 E_{\mathrm{res}} \simeq 13200 \, \cos 2 \theta_{13} \, \frac{\ldm \, [\mathrm{eV^2}]}{\rho \, [\mathrm{g \, cm^{-3}}]}  \label{equ:msw}
\end{equation}
for neutrinos and the normal mass ordering, or antineutrinos and the inverted mass ordering.  For the average density of the Earth's mantle and small $\stheta$, one finds $ E_{\mathrm{res}}  \simeq 7 \, \mathrm{GeV}$, while  the average density of the Earth's (outer) core gives $E_{\mathrm{res}}  \simeq 3 \, \mathrm{GeV}$.

\begin{figure}
\begin{center}
\includegraphics[width=0.49\textwidth]{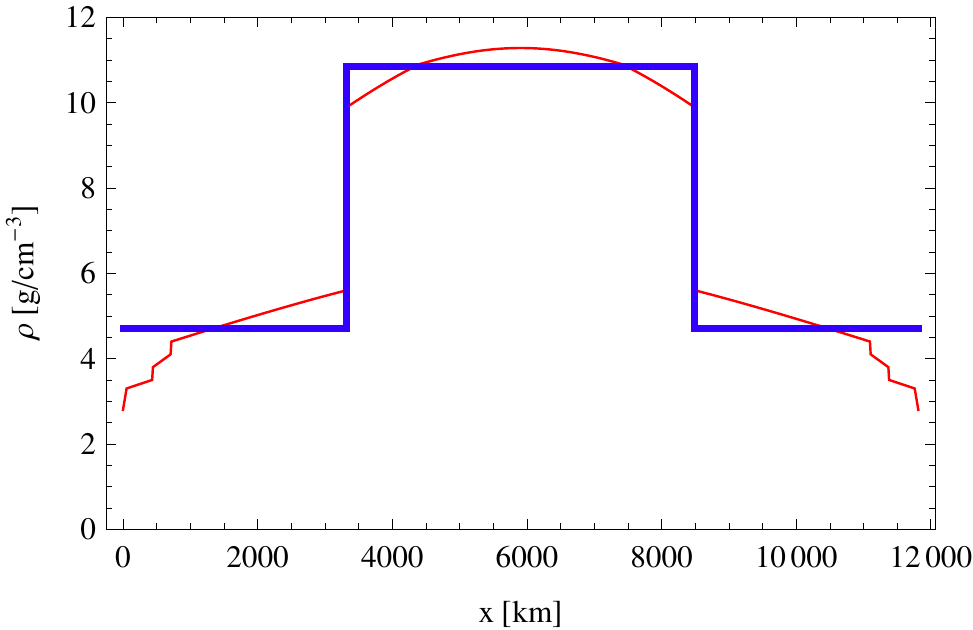} \includegraphics[width=0.49\textwidth]{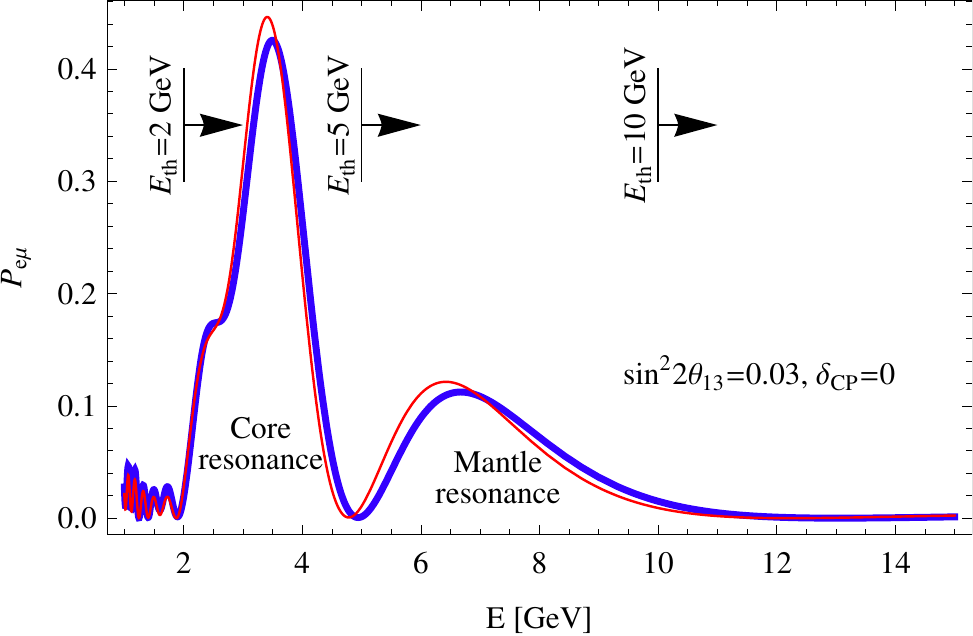}
\end{center}
\mycaption{\label{fig:profile} Left panel: Matter density profile for a baseline $L=11810 \, \mathrm{km}$ for the Reference Earth Model~\cite{Dziewonski:1981xy} (thin curve) and a three-step approximation (thick curve). Right panel: $P_{e \mu}$ appearance probability for these two matter density profiles (same line styles; normal hierarchy).}
\end{figure}

Since the major high energy accelerator laboratories are in the northern hemisphere, some potential baselines to \dc\ will cross the Earth's core. More specifically, some baselines are\footnote{See discussion in \Ref~\cite{Agarwalla:2010hk} for a neutrino factory.}
\begin{itemize}
\item
CERN-\dc : $11810 \, \mathrm{km}$
\item
FNAL-\dc : $11620 \, \mathrm{km}$
\item
JPARC-\dc : $11370 \, \mathrm{km}$
\item
RAL-\dc : $12020 \, \mathrm{km}$
\end{itemize}
Considering oscillation physics, these baselines are very similar and the physics outcome will be the same. Therefore CERN-\dc : $11810 \, \mathrm{km}$ is chosen as an example for this 
study. We show in \figu{profile}, left panel, the matter density profile for  $L=11810 \, \mathrm{km}$ for the Reference Earth Model (thin curve), and a  three-step approximation (thick curve) consisting of the average densities in the qualitatively different layers. Compared to \equ{papp}, there are two very different matter densities present at the same time, leading to a ``parametric enhancement''~\cite{Akhmedov:1998ui,Chizhov:1998ug}. We show in the right panel the oscillation probability $P_{e \mu}$, which takes relatively large values because of this effect. Here the different curves correspond to the matter density profiles in the left panel. Obviously, the two curves are qualitatively similar, which means that the three step profile is a reasonable  approximation that we apply. Note that the parametric resonance cannot be correctly reproduced with a single density approximation, leading to qualitatively incorrect conclusions. In the right panel the two peaks from the core resonance and mantle resonance can also be seen. From this figure we may extract  that a detector threshold as low as 2~GeV will be needed to cover the core resonance, and a detector threshold as low as 5~GeV to cover the mantle resonance. For a threshold much lower than 2~GeV, no qualitatively improvement is expected, whereas for a threshold significantly above 10~GeV, no particular sensitivity from the appearance channel is expected.

\begin{figure}
\begin{center}
\includegraphics[width=0.8\textwidth]{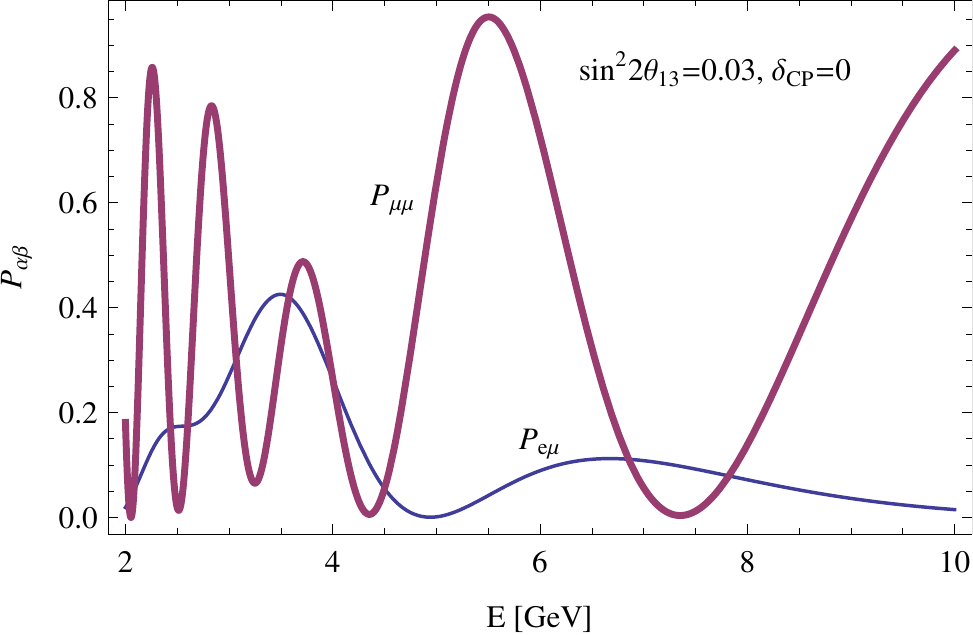}
\end{center}
\mycaption{\label{fig:probcomp} Comparison between the survival $P_{\mu \mu}$ and appearance $P_{e \mu}$ probabilities as a function of energy for $L=11810 \, \mathrm{km}$ and $\stheta=0.03$, $\deltacp=0$. }
\end{figure}

Having discussed the appearance probability above, let us now focus on the survival probability. 
First, in \figu{probcomp}, we show a comparison between the survival $P_{\mu \mu}$ and appearance $P_{e \mu}$ probabilities. It is noteworthy that, for large enough values of $\stheta$, there are energy windows where the appearance probability dominates, and others where the survival probability dominates. This effect will be relevant for a neutrino factory if no charge identification can be used in the detector~\cite{Huber:2008yx} and the appearance and survival probabilities cannot be distinguished. One can easily see from \figu{probcomp} that, with sufficient energy resolution, the signals from these two probabilities can be separated. Will will quantify in \Sec~\ref{sec:reqnf} the needed  energy resolution  for this separation.

\begin{figure}
\begin{center}
\includegraphics[width=\textwidth]{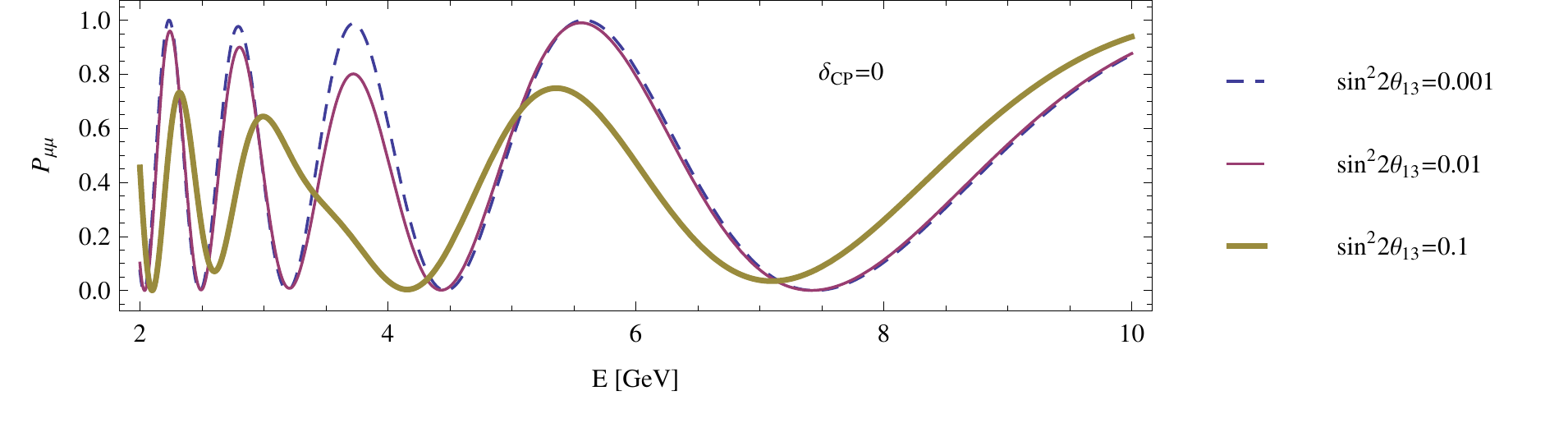}

\includegraphics[width=\textwidth]{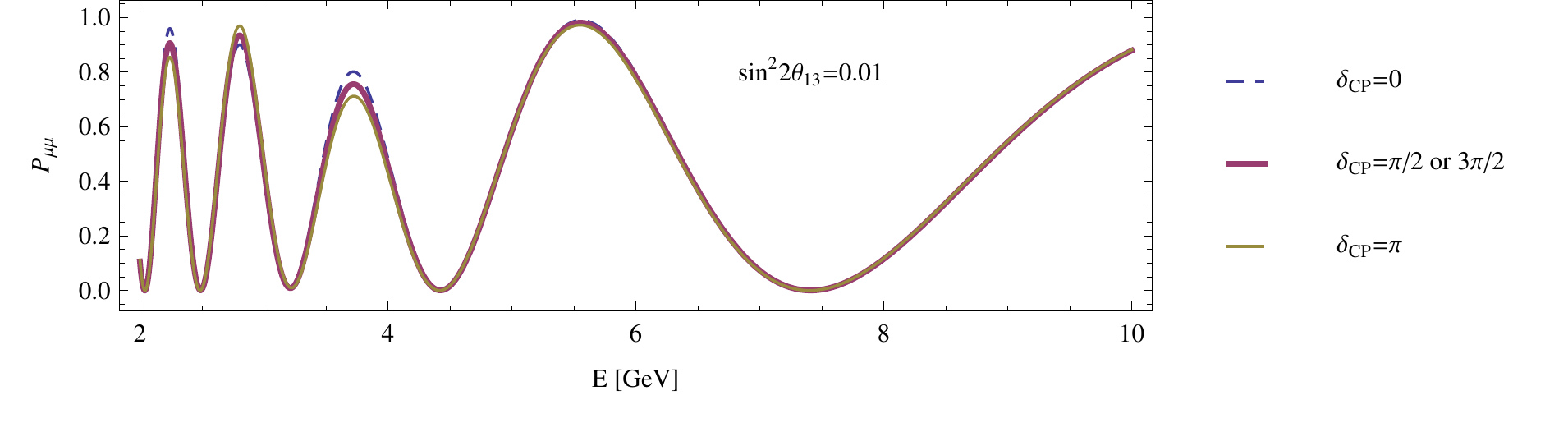}
\end{center}
\mycaption{\label{fig:disapp}Disappearance probability ($P_{\mu \mu}$) as a function of energy for different values of $\stheta$ (upper panel) and different values of $\deltacp$ (lower panel).}
\end{figure}

For large values of $\stheta$, not only the appearance probability contains the information on $\deltacp$, but also the survival probability $P_{\mu \mu}$ contributes~\cite{Akhmedov:2004ny}. While a survival probability can, in general, not be directly sensitive to the CP violating term $\propto \sin \deltacp$, the dependence on $\cos \deltacp$ is maintained. This may be used to discriminate CP conserving ($\cos \deltacp=1$) from CP violating ($\cos \deltacp \sim 0$) solutions. Nevertheless, high statistics are needed to establish this signal against the background from the leading (atmospheric) oscillation. We show in \figu{disapp} the survival probability as a function of energy for different values of $\stheta$ (upper panel) and different values of $\deltacp$ (lower panel). While $\stheta$ affects the amplitude of all oscillation maxima in the same way, a different value of $\deltacp$, compared to the maximally CP violating values $\pi/2$ or $3 \pi/2$, leads to a modulation of the oscillation maxima in different directions. In principle, $\stheta$ and $\deltacp$ can be disentangled by measuring more than one oscillation maximum. Again, energy resolution is important for this effect. As indicated above, the modulation of the signal is relatively small, meaning that very high statistics are required and therefore a large detector may help. 

There is other oscillation physics within the neutrino Standard Model that can be performed at very long or core-crossing baselines beyond the $\stheta$, $\deltacp$, and the mass hierarchy discoveries. For example, such a baseline is helpful (compared to, or combined with a shorter baseline) for a measurement of the atmospheric oscillation parameters~\cite{Huber:2006wb}, deviations from maximal atmospheric mixing~\cite{Huber:2006wb}, determination of the $\theta_{23}$ octant~\cite{Gandhi:2006gu}, and $\deltacp$~\cite{Huber:2004gg} and $\theta_{13}$~\cite{Gandhi:2006gu} precision measurements. Furthermore, the MSW effect in Earth matter may be verified at a high confidence level, even for $\stheta=0$, by the solar appearance term (the fourth term in \equ{papp})~\cite{Winter:2004mt}. Apart from standard oscillation physics, certain classes of new physics effects are best tested at such a long baseline, including  ``non-standard interactions'' (see Fig.~6 in \Ref~\cite{Kopp:2008ds}). One interesting geophysical application, which we discuss  \Sec~\ref{sec:earth}, is the determination of the Earth's matter density along the baseline, including that which is core-crossing, via the MSW effect~\cite{Winter:2005we,Gandhi:2006gu,Minakata:2006am}. This application is unique for the setups used in this study, since a baseline to the South Pole is the only currently investigated option for a core-crossing baseline using a neutrino beam.

\section{Reference setups}
\label{sec:reference}

\begin{table}[t]
\begin{center}
\begin{tabular}{lllrlr}
\hline
Label & Beam & Detector~1 & $L_1$ [km] & Detector~2 & $L_2$ [km] \\ 
\hline
\multicolumn{6}{l}{\bf{Two-baseline setups for small $\boldsymbol{\theta_{13}}$}:} \\
BB$_{\mathrm{ref}}$ & Beta beam (four isotopes) & WC & 650 & MIND & 7000  \\
NF$_{\mathrm{ref}}$ & Neutrino factory, $E_\mu=25 \, \mathrm{GeV}$ & MIND & 7500 & MIND & 4000 \\
BB$_{\mathrm{bm}}$ & Beta beam (four isotopes) & WC & 650 & \dc\ & 11810  \\
NF$_{\mathrm{bm}}$ & Neutrino factory, $E_\mu=25 \, \mathrm{GeV}$ & MIND & 7500 & \dc\ & 11810 \\
\hline
\multicolumn{6}{l}{\bf{Single-baseline setups for large $\boldsymbol{\theta_{13}}$}:} \\
BB$_{\mathrm{ref}}^*$ & Beta beam ($^{18}$Ne and $^6$He) & WC & 650 &  \multicolumn{2}{c}{n/a}  \\
NF$_{\mathrm{ref}}^*$ & Neutrino factory, $E_\mu=10 \, \mathrm{GeV}$ & MIND & 2200 &  \multicolumn{2}{c}{n/a} \\
SB$_{\mathrm{ref}}^*$ & LBNE experiment & WC & 1300 &  \multicolumn{2}{c}{n/a}\\
BB$_{\mathrm{bm}}^*$ & Beta beam ($^8$B  and $^8$Li) & \dc\ & 11810 & \multicolumn{2}{c}{n/a}  \\
NF$_{\mathrm{bm}}^*$ & Neutrino factory, $E_\mu=5 \, \mathrm{GeV}$ & \dc\ & 11810&  \multicolumn{2}{c}{n/a} \\
SB$_{\mathrm{bm}}^*$ & Superbeam (LBNE-like) & \dc\ & 11810 &  \multicolumn{2}{c}{n/a} \\
\hline
\end{tabular}
\end{center}
\mycaption{\label{tab:setups} Description and terminology of the reference (subscript ``ref'') and benchmark (subscript ``bm'') setups used in this work.}
\end{table}

Here we introduce our reference setups for beta beam, neutrino factory, and superbeam, as well as the beam technologies and detector requirements. These reference setups will be used later to compare with configurations including the \dc\ detector. Note that for beta beam and neutrino factory, we define configurations with two baselines (small $\theta_{13}$) and one baseline (large $\theta_{13}$). 
In the following, the two-baseline reference setups will be referred to as BB$_{\text{ref}}$ (beta beam) and NF$_{\text{ref}}$ (neutrino factory), the single-baseline setups as BB$_{\text{ref}}^*$ (beta beam), NF$_{\text{ref}}^*$ (neutrino factory), and SB$_{\text{ref}}^*$ (superbeam). The reference setups (subscript ``ref'') are summarized in \Tab~\ref{tab:setups}.

In all of the following, we perform the numerical simulations using the  GLoBES software package~\cite{Huber:2004ka,Huber:2007ji}.  
The best-fit oscillation parameters are taken as follows~\cite{GonzalezGarcia:2010er}:
\begin{align}
&\theta_{12}=33.8^\circ\,,\quad\theta_{13}=5.6^\circ\,,\quad\theta_{23}=45^\circ
\nonumber\\
&\Delta m_{21}^2=7.7\times10^{-5}\text{eV}^2\,,\quad\Delta m_{31}^2=2.4\times10^{-3}\text{eV}^2\,.
\end{align}
We impose external $1\sigma$ errors on $\Delta m^2_{21}$ (1\%) and
$\theta_{12}$ (1\%) and on $\Delta m^2_{31}$ (5\%) and
$\theta_{23}$ (5\%) as conservative estimates for the current
measurement uncertainties~\cite{GonzalezGarcia:2010er}. 
We also include a 5\% matter density uncertainty~\cite{Geller:2001ix,Ohlsson:2003ip} in the standard performance indicators. Typically, we show the results for the simulated normal hierarchy. Note, however, that because of the symmetric operation with neutrinos and antineutrinos, the difference for the inverted hierarchy is small.\footnote{The only exception may be the beta beam, where the $\bar\nu_e$ from $^8$Li have lower energies than the $\nu_e$ from $^8$B, see below. Here the normal hierarchy performance should be somewhat better.}

\subsection{Beta beam references (BB$_{\text{ref}}$, BB$_{\text{ref}}^*$)}

Currently discussed configurations of a beta beam produce a neutrino beam by beta decays  of the following proposed isotopes:
\begin{eqnarray}
_{10}^{18}\text{Ne} & \rightarrow & _{9}^{18}\text{F} + e^+ + \nu_e  \, ,\\
_{4}^{6}\text{He} & \rightarrow  & _{3}^{6}\text{Li} + e^- + \bar{\nu}_e  \, ,\\ 
_{5}^{8}\text{B} & \rightarrow & _{4}^{8}\text{Be} + e^+ + \nu_e \, \\
_{3}^{8}\text{Li} & \rightarrow  & _{4}^{8}\text{Be} + e^- + \bar{\nu}_e  \, ,
\end{eqnarray}
stored in a storage ring with long straight sections (see Fig.~3 in \Ref~\cite{Tang:2009na} for such a storage ring). The oscillation channels of interest are:
\begin{align}
\nu_\mu \text{ appearance:   } & \nu_e \rightarrow \nu_\mu \text{ for } ^{18}\text{Ne} \text{ or } ^{8}\text{B}\text{ stored} \, , \\
\bar\nu_\mu \text{ appearance:   } & \bar\nu_e \rightarrow \bar\nu_\mu \text{ for } ^{18}\text{He} \text{ or } ^{8}\text{Li}\text{ stored} \, .
\end{align}
Since the beam is flavor-clean, there are no special detector requirements other than $\nu_\mu$ ($\bar \nu_\mu$) flavor identification. This, a beta beam is {\em per se} the best candidate for \dc . 

The isotopes  $^8$B and $^8$Li have higher endpoint energies $E_0$ than  $^{18}$Ne and $^6$He, which means that the same neutrino energies $E_\nu \propto \gamma E_0$ can be reached with lower boost factors $\gamma$, and therefore less acceleration effort. In turn, higher neutrino energies can be reached with the same accelerator. Note that the total flux in the detector is proportional to $\gamma^2$, meaning that higher boost factors produce higher event rates. Since beta beams are typically discussed in context with the CERN-SPS as accelerator, possibly including upgrades, the maximum $\gamma$ is limited. As a consequence, for small $\theta_{13}$, $^{18}$Ne and $^6$He are proposed for CP violation measurements at shorter baselines, whereas $^8$B and $^8$Li may be used at a longer baseline for determining the mass hierarchy; see \Refs~\cite{Agarwalla:2008gf,Choubey:2009ks} for the optimization.  The chosen detector technology for a beta beam typically depends on the neutrino energies: below 1~GeV a good separation between the quasi-elastic event from electron and muon neutrinos and an efficient background reduction can be reached in a WC detector, a (possibly magnetized) iron calorimeter can better identify muon tracks at energies above 1~GeV.

We use the four-ion setup~\cite{Choubey:2009ks} for reference BB$_{\text{ref}}$, which we have reproduced using the detector simulation~\cite{Huber:2005jk,BurguetCastell:2005pa}. In this setup, \neon and \he are each accelerated to $\gamma = 350$ and detected in a 500 kton WC  
 detector located at Canfranc in Spain with a baseline of 650~km. \br and \li are accelerated to $\gamma = 656$  and $\gamma = 390$, respectively, and detected at 7000 km by a 50 kton magnetized iron
 detector (MIND), possibly at INO in India. 
We use a total running time of ten years with $10^{19}$ useful ion decays per year in total, effectively $2.5$ years per each ion species. In practice, the luminosity may be lower, but distributed over a longer time, since some of the modes can be run in parallel. This number is ambitious and yet to be proven, potentially using the production ring technology~\cite{euronu}. The signal normalization errors are 2.5\%, with background normalization errors of 5\%.
The peak neutrino energies in this setup are roughly 1.4~GeV for $^{18}$Ne and $^6$He stored, 9.5~GeV for $^8$B, and 5.2~GeV for $^8$Li. We assume that the straights of the storage ring for the very long baseline are only 60\% of that of the short baseline storage ring~\cite{Choubey:2009ks} due to the steepness, and  this yields $8 \cdot 10^{19}$ useful ion decays in total. 
For large $\theta_{13}$, we use a similar setup BB$_{\text{ref}}^*$ with the short baseline only, which is close to optimal~\cite{Winter:2008cn} and also considered within the Euronu design study~\cite{euronu}. 
The total number of useful $^8$B and $^8$Li decays is the same as for BB$_{\text{ref}}$, and no additional technology for $^{18}$Ne and $^6$He is needed, yielding $3 \cdot 10^{19}$ useful ion decays in total. 

\begin{figure}[t]
 \centering
 \includegraphics[width=0.9\textwidth]{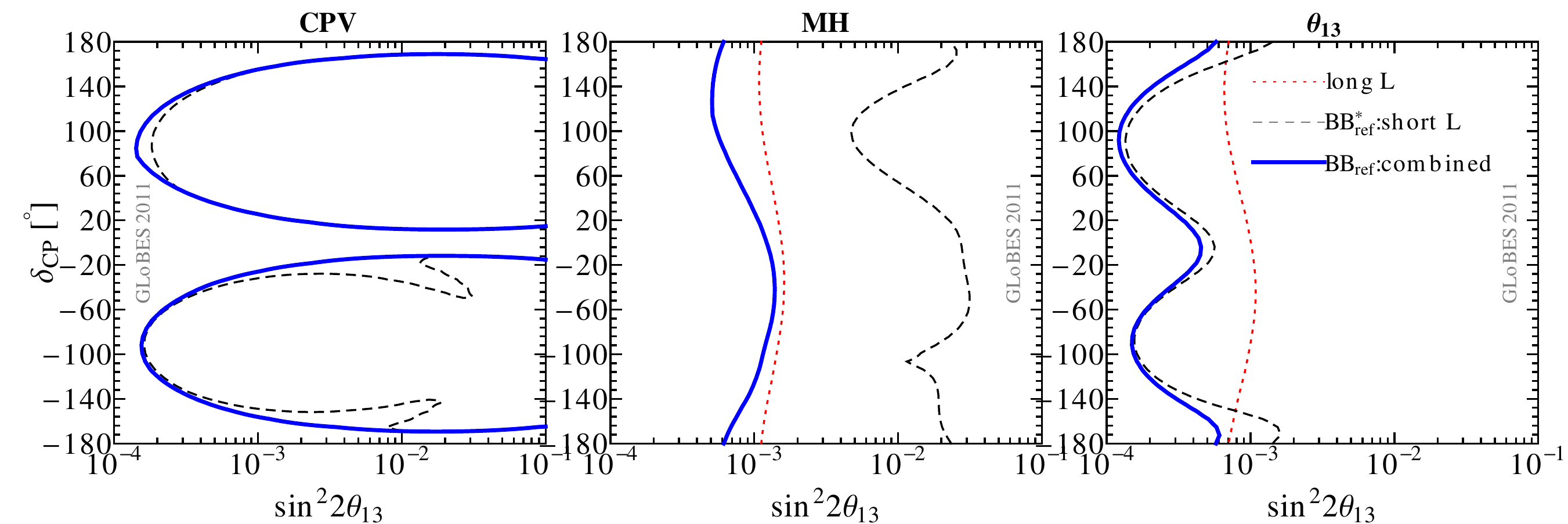}
 \includegraphics[width=0.9\textwidth]{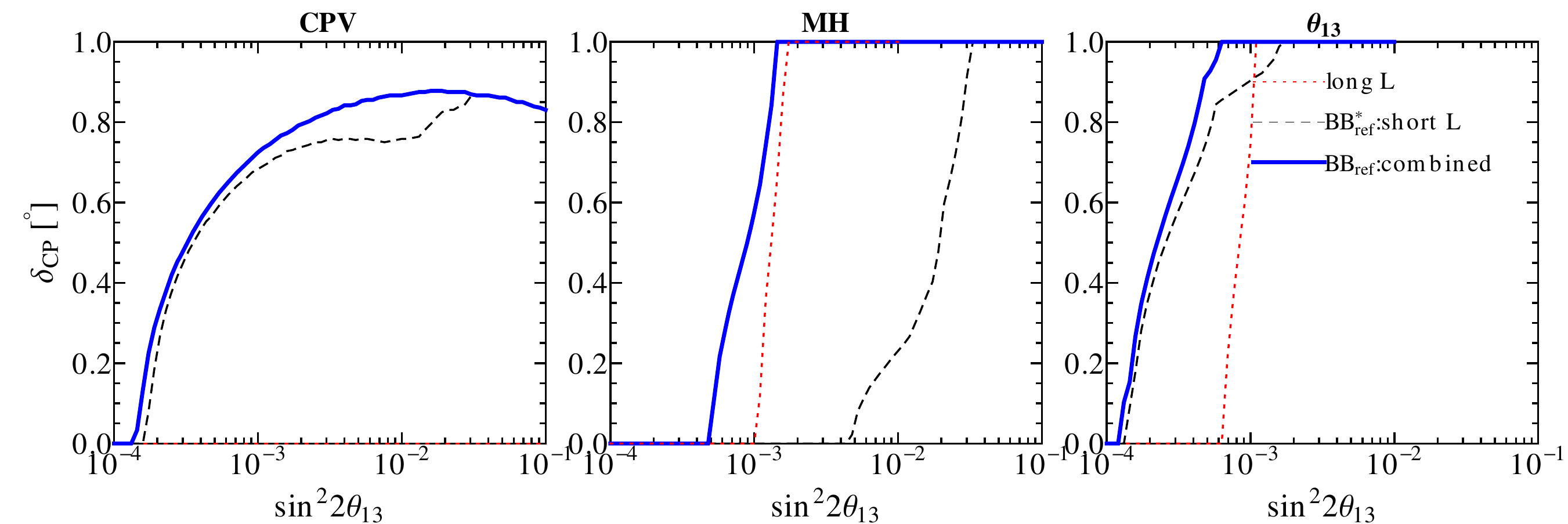}
 \mycaption{Discovery reach for CPV, MH and $\theta_{13}$ at the $3\sigma$ confidence level for the reference beta beam setups BB$_{\text{ref}}$ and BB$_{\text{ref}}^*$, where the short and long baseline reaches are shown separately. Note that the short baseline of BB$_{\text{ref}}$ corresponds to BB$_{\text{ref}}^*$. Here we assume the normal hierarchy.}
 \label{fig:BB-ref}
\end{figure}

Shown in \figu{BB-ref} is the performance of the reference beta beam setups as a function of true $\stheta$ and $\deltacp$ (upper row) or fraction of $\deltacp$ (lower row) for the CPV, MH, and $\theta_{13}$ discovery reaches. In these panels, the parameter of interest will be discovered on the r.h.s. of the curve. The thick curves show the two-baseline setup BB$_{\text{ref}}$, which have excellent discovery reaches for small values of $\stheta$. The difference between the one- and two-baseline setups can be seen by comparing  the dashed and solid curves: the second (longer) baseline especially aids degeneracy resolution for small $\theta_{13}$ (left panels) and for the mass hierarchy determination for small $\theta_{13}$ (middle panels), whereas it may not be required for large $\theta_{13}$. The large $\theta_{13}$ reference setup BB$_{\text{ref}}^*$ (with only one baseline) corresponds to the dashed curves.
Later we will either replace the long baseline in the two-baseline setup by a baseline to \dc , or the single (short) baseline for large $\theta_{13}$, and compare the performance to these reference setups.

\subsection{Neutrino factory references (\NFref, \NFrefs)}

The neutrino factory beam comes from muon decays (muon energy $E_\mu$) in straight sections of a storage ring. Depending on the parent muon, the decay will lead to
\begin{eqnarray}
+ & : & \mu^- \rightarrow e^- + \bar\nu_e + \nu_\mu \, , \\
- & : & \mu^+ \rightarrow e^+ + \nu_e + \bar\nu_\mu  \, ,
\end{eqnarray}
where we have defined the polarities ``+'' and ``--''.
This implies that for any given muon polarity, two different neutrino flavors, one neutrino and one antineutrino, are produced at the same time. The oscillation channels of interest are:
\begin{align}
\nu_\mu \text{ appearance:   } & \nu_e \rightarrow \nu_\mu \text{ for } \mu^+ \text{ stored} \, \, (-) \, , \label{equ:numu} \\
\bar\nu_\mu \text{ disappearance:   } & \bar\nu_\mu \rightarrow \bar\nu_\mu \text{ for } \mu^+ \text{ stored} \label{equ:numubar} \, \, (-) \, , \\
\bar\nu_\mu \text{ appearance:   } & \bar\nu_e \rightarrow \bar\nu_\mu \text{ for } \mu^- \text{ stored} \, \, (+) \, ,\\
\nu_\mu \text{ disappearance:   } & \nu_\mu \rightarrow \nu_\mu \text{ for } \mu^- \text{ stored} \, \,  (+) \, . 
\end{align}
Since the different muon polarities can be distinguished by the beam bunching, the neutrinos from $\nu_\mu$ ($\bar \nu_\mu$) appearance and $\bar\nu_\mu$ ($\nu_\mu$) disappearance will be present at the same time in the detector, and have to be distinguished. This identification is typically performed with a magnetic field that permits  the charge identification of the muon produced in a charged current interaction. The typical choice for the detector technology is therefore a large magnetized iron detector (MIND). Since charge identification is probably unrealistic in \dc , the most practicable other methods to discriminate between the channels may be energy resolution~\cite{Huber:2008yx}; see discussion in \Sec~\ref{sec:oscillation}. This, however, is potentially also difficult in a detector such as \dc .

For a neutrino factory, the small and large $\theta_{13}$ setups are specified in the IDS-NF Interim Design Report~\cite{IDSNF:2011aa}. For small values of $\theta_{13}$, a two-baseline high energy neutrino factory (HENF) setup (referred to by \NFref) with a 100~kt MIND at $L_1 \simeq 3000$--$5000 \, \mathrm{km}$ and a 50~kt MIND at $L_2 \simeq 7500 \, \mathrm{km}$ is proposed, muon energy $E_\mu = 25 \, \mathrm{GeV}$, recently optimized in~\cite{Huber:2006wb,Kopp:2008ds,Agarwalla:2010hk}. For large values of $\theta_{13}$, a single-baseline setup (referred to by \NFrefs) with a smaller $E_\mu$ and shorter baselines is preferable, as proposed as low energy neutrino factory (LENF) in \Refs~\cite{Geer:2007kn,Bross:2007ts,Bross:2009gk}. We use the recent data on the MIND detector~\cite{Abe:2007bi,Cervera:2010rz,ThesisLaing} for our simulation. This leads to an optimized LENF reference setup with $E_\mu=10 \, \mathrm{GeV}$ and $L=2200 \, \mathrm{km}$~\cite{Agarwalla:2010hk} for which we use the 100~kt MIND detector. We also include the $\nu_\tau$ contamination from the decay mode of tau leptons (produced by $\nu_\tau$) into muons in the detector~\cite{Indumathi:2009hg,Donini:2010xk}, including the corresponding migration matrices~\cite{Agarwalla:2010hk}. Since this background is typically reconstructed at lower energies, and may fall into the windows where the appearance information dominates,  this contamination may be potentially harmful if energy resolution is used for the channel discrimination in the detector. We consider the impact of this effect later in the paper. We assume a total $10^{21}$ useful muon decays per year, with a total operation time of ten years. This number of useful muon decays must be divided among four decay straights (two rings with two polarities each) for the HENF, and among two decay straights for the LENF. 
The signal normalization errors are 2.5\%, with background normalization errors of 20\%. The peak neutrino energies are about 25~GeV for \NFref\ and 10~GeV for \NFrefs\ for muon neutrinos, and 17~GeV for \NFref\ and 8~GeV for \NFrefs\ for electron neutrinos.

\begin{figure}[t]
 \centering
 \includegraphics[width=0.9\textwidth]{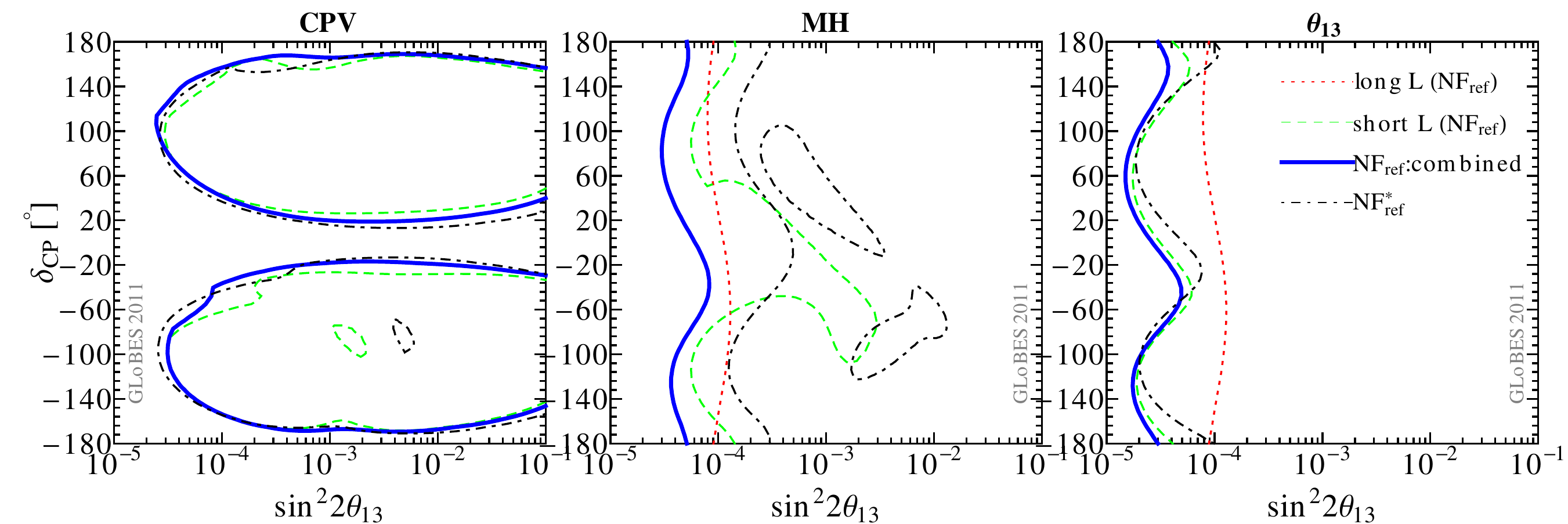}
 \includegraphics[width=0.9\textwidth]{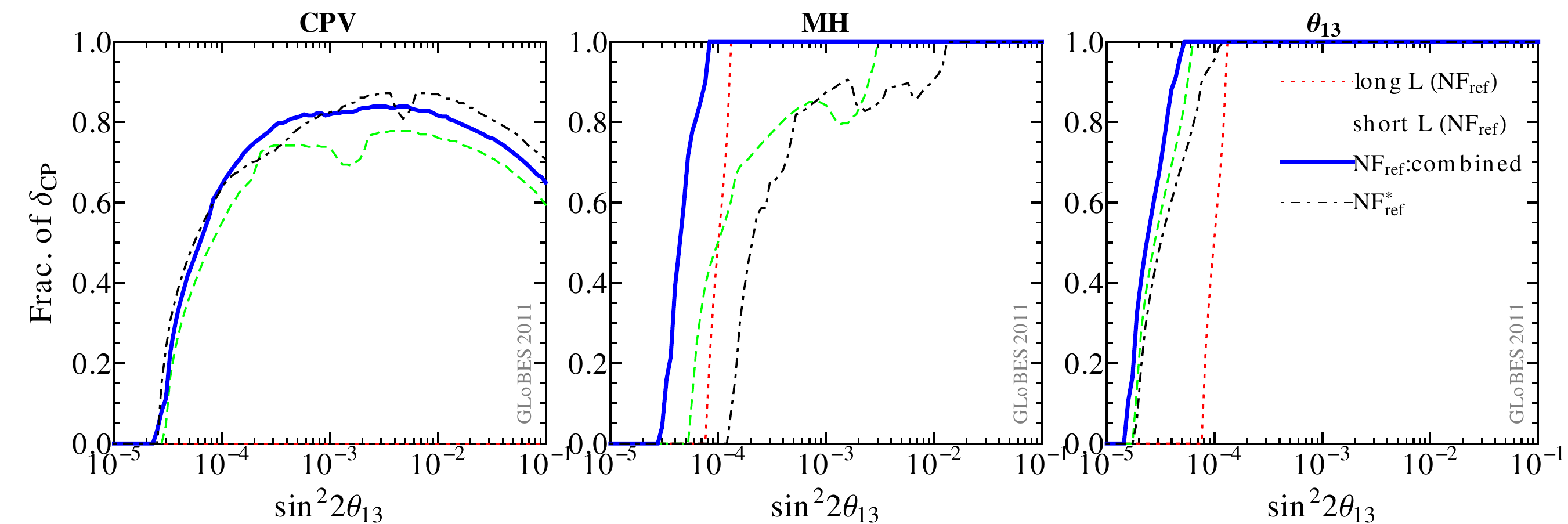}
 \mycaption{Discovery reach to CPV, MH, and $\theta_{13}$ at $3\sigma$ for the reference neutrino factory setups \NFref\ and \NFrefs, where the short and long baseline reaches (\NFref) are shown separately. Here we assume the normal hierarchy. }
 \label{fig:NF-ref}
\end{figure}

Shown in  \figu{NF-ref} is the performance of the reference setups. For \NFref , the main performance for CPV comes from the short baseline, but the long baseline assists degeneracy resolution. For the mass hierarchy resolution, the long baseline dominates the performance. For $\theta_{13}$, the short baseline performs better, primarily due to the larger detector mass placed there. Later, we will replace the long baseline detector by the \dc\ detector. For \NFrefs , all performance indicators can be measured when $\theta_{13}$ is large enough, whereas for small $\theta_{13}$ the usual degeneracy problems are encountered. In addition, the CPV performance for large $\theta_{13}$ is better than for \NFref . We will compare the \dc\ setup for large $\theta_{13}$ later to this reference setup.

\subsection{Superbeam reference (\SBrefs)}

Conventional neutrino beams, such as T2K and MINOS, are produced  mainly from the decay of pions produced in  interactions of a proton beam with a fixed target:
\begin{eqnarray}
 \pi^+\to&\mu^++\nu_\mu \, ,\\
\pi^-\to&\mu^-+\bar{\nu}_\mu \, .
\end{eqnarray}
Since the muons produced (and also kaons) will further decay, a non-negligible background of $\nu_e$ and $\bar\nu_e$ is unavoidable. This leads to a systematics limitation for small values of $\stheta$, which means that superbeams typically have a limited reach in $\stheta$.
It is possible to use the off-axis technology~\cite{offaxis} to produce a narrow band beam and to over-proportionally reduce the backgrounds, such as in T2K and NO$\nu$A.  In MINOS and LBNE, the beam is directly aimed towards the detector, leading to a wide neutrino spectrum. In the LBNE case, the spectrum peaks near 2.5 GeV  which means that this beam has the ideal energies for our purpose.
The leading oscillation channels of interest are
\begin{align}
\nu_e \text{ appearance:   } & \nu_\mu \rightarrow \nu_e \text{ for } \pi^+ \text{ decays} \, , \\
\nu_\mu \text{ disappearance:   } & \nu_\mu \rightarrow \nu_\mu \text{ for } \pi^+ \text{ decays} \, ,\\
\bar{\nu}_e \text{ appearance:   } & \bar{\nu}_\mu \rightarrow \bar{\nu}_e \text{ for } \pi^- \text{ decays} \, , \\
\bar\nu_\mu \text{ disappearance:   } & \bar\nu_\mu \rightarrow \bar\nu_\mu \text{ for } \pi^- \text{ decays} \, . 
\end{align}
Thus, for a superbeam detector no particular requirements other than flavor identification are necessary, and WC detectors are often proposed. Note that compared to the beta beam, the electron flavor must be identified for the appearance channel, which may be more challenging than the muon flavor in a detector like \dc .
In either case, since the neutrinos carry only a small fraction of the initial proton energy, the intensity of conventional beams is rather limited even for high thermal target powers. It therefore remains to be seen if such a beam has enough intensity to power a core-crossing baseline.

\begin{figure}[!t]
 \centering
 \includegraphics[width=0.9\textwidth]{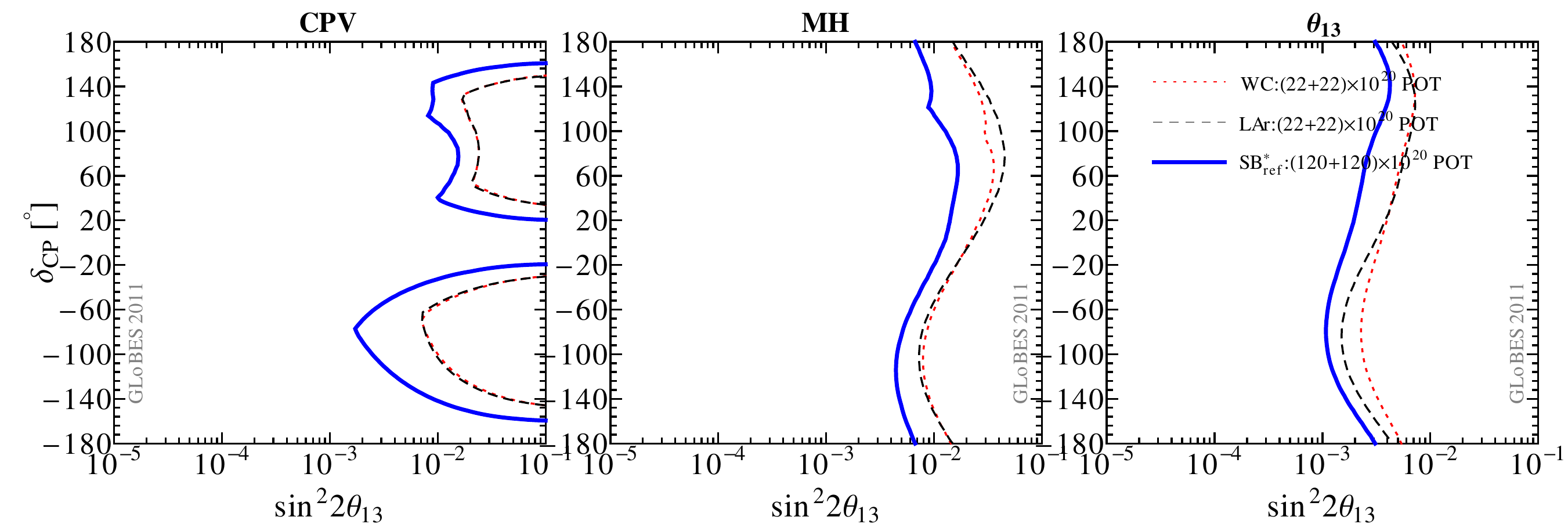}
 \includegraphics[width=0.9\textwidth]{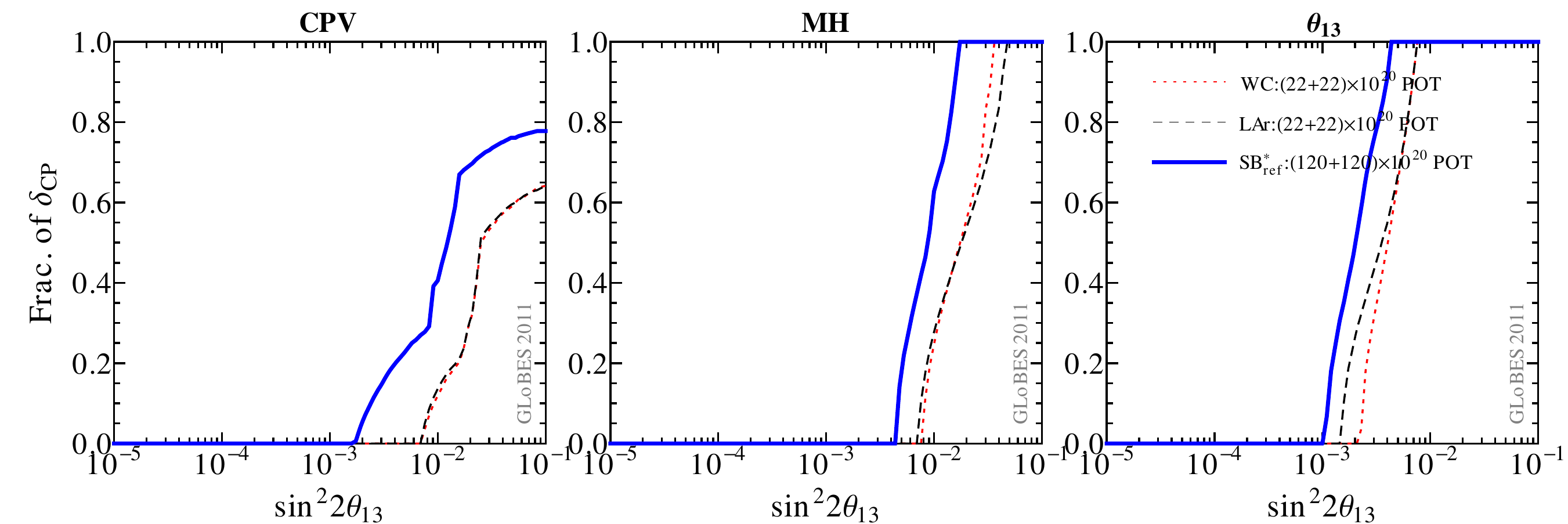}
 \mycaption{A comparison of the discovery reach of CPV, MH and $\theta_{13}$ at $3\sigma$ for LBNE with different luminosities and detectors, including the reference \SBrefs . Here we assume the normal hierarchy. }
 \label{fig:LBNE-ref}
\end{figure}

We use the 120~GeV configuration of the US-based LBNE project~\cite{LBNE} as a prototype, referred to as \SBrefs .  
There were two candidates for the far detector: one is a 200 kton fiducial mass WC detector; the other is a 34 kton Liquid Argon detector. The baseline is 1300 km from FNAL to the Homestake mine in South Dakota. We choose LBNE with a WC detector as our reference setup for a superbeam experiment, with a total of $120 \cdot 10^{21}$ (neutrino mode) plus $120 \cdot 10^{21}$ (antineutrino mode) protons on target. This corresponds to five years of operation in each mode with 2.3~MW target power.  For the beam and detector simulation, we use the information from the official webpage~\cite{LBNE}, and cross checked with~\cite{Huber:2010dx}. The signal and background normalization errors are $5\%$ and $10\%$, respectively, for the disappearance channels, and  $1\%$ and $5\%$, respectively, for the appearance channels. We show in \figu{LBNE-ref} the performance of this reference setup, compared to different detector and luminosity configurations. The optimal CPV and MH discovery reaches can only be obtained for $\stheta \gtrsim 0.01$ and therefore we only consider this single-baseline setup for large $\theta_{13}$. Later we will compare it to a corresponding \dc\ configuration.

\section{Parameterization of the  \dc\ detector}
\label{sec:detector}

Compared to traditional WC detectors, where the photodetectors are placed at the surface of a tank, the photodetectors in a conventional neutrino telescope are attached to strings distributed within the detection volume. This means that three dimensional space and timing information can be used for the event reconstruction, which implies that the event topologies in a neutrino telescope are somewhat different from a WC detector~\cite{Beacom:2003nh}. An extraterrestrial neutrino detector typically utilizes muon tracks, which are most dominantly produced by charged current interactions of muon neutrinos. Other event topologies include electromagnetic (from electron neutrinos), hadronic (from tau neutrinos), and neutral current cascades. Since it is relatively difficult to distinguish among the different types of cascades, one has to accept either large contaminations (flavor-misidentification) for electron and tau neutrinos, or apply significant cuts to identify events for which the flavor can be uniquely assigned. In this work, we assume the same fiducial mass for all flavors as a first approximation. This is  a reasonable assumption from preliminary fiducial mass computations~\cite{Koskinen}, and we parameterize the mis-identification rate of the events. In future simulations, the optimal  balance between cuts and mis-identification has to be determined.\footnote{Stronger cuts typically imply a lower mis-identification rate coming together with a smaller fiducial mass, and therefore lower statistics. A priori it is not clear where the optimum balance between fiducial volume (statistics) and mis-identification (systematics) lies in terms of the physics performance, which requires some optimization of the cuts.} 
Other possible event topologies include the possibility of a separated tau track at PeV energies~\cite{Learned:1994wg}.
The exact computation of event rates (or fiducial masses) for different topologies requires a detailed detector simulation of the \dc\ detector. 

\begin{table}[t!]
\begin{center}
\begin{tabular}{llll}
\hline
Setup(s) & Signal & Factor & Background \\ 
\hline
\BBbm , \BBbms & $\nu_e \rightarrow \nu_\mu$ & misID & $\nu_e \rightarrow \nu_x$, neutral current (cascades as tracks) \\
               &  & misID & $\nu_e \rightarrow \nu_e$, CC, mis-ID (cascades as tracks) \\
\hline
\NFbm , \NFbms & $\bar\nu_e \rightarrow \bar\nu_\mu$  & misID & $\bar\nu_e \rightarrow \bar\nu_x$, neutral current (cascades as tracks) \\
&  $\nu_\mu \rightarrow \nu_\mu$ & misID & $\nu_\mu \rightarrow \nu_x$, neutral current (cascades as tracks) \\
&  & misID & $\bar\nu_e \rightarrow \bar\nu_e$, CC,  mis-ID (cascades as tracks) \\
&  & misID & $\nu_\mu \rightarrow \nu_e$, CC,  mis-ID (cascades as tracks) \\
&  & misID & $\bar\nu_e \rightarrow \bar\nu_\tau$, CC,  mis-ID (cascades as tracks) \\
&  & misID & $\nu_\mu \rightarrow \nu_\tau$, CC,  mis-ID (cascades as tracks) \\
& & 17\% & $\bar\nu_e  \rightarrow \bar\nu_\tau \rightarrow \tau^+ \overset{17\%}{\rightarrow} \mu^+$ (intrinsic $\tau$ cont.) \\
& & 17\% & $\nu_\mu  \rightarrow \nu_\tau \rightarrow \tau^- \overset{17\%}{\rightarrow} \mu^-$ (intrinsic $\tau$ cont.) \\ 
\hline
\SBbms & $\nu_\mu \rightarrow \nu_e$ & 1.0 & $\nu_e \rightarrow \nu_e$, intrinsic beam background \\
 & & misID & $\nu_\mu \rightarrow \nu_x$, neutral current (cascades as cascades) \\
 & & misIDtracks & $\nu_\mu \rightarrow \nu_\mu$, CC, mis-ID (tracks as cascades) \\
 & & misID & $\nu_\mu \rightarrow \nu_\tau$, CC, mis-ID (cascades as cascades) \\
& $\nu_\mu \rightarrow \nu_\mu$ & misIDtracks & $\nu_\mu \rightarrow \nu_e$, CC, mis-ID (cascades as tracks) \\
& &  misIDtracks & $\nu_e \rightarrow \nu_e$, CC, mis-ID (cascades as tracks) \\
& &  misIDtracks & $\nu_\mu \rightarrow \nu_x$, neutral current (cascades as tracks) \\
\hline
\end{tabular}
\end{center}
\mycaption{\label{tab:backgrounds} Signal and backgrounds used in the parameterization of the \dc\ detector. In the GLoBES software, the oscillated background rates are calculated, multiplied by ``Factor'' (and the background systematical error parameter), and then added to the signal event rate.
Here $\nu_x$ refers to all flavors, and ``CC'' to charged current interactions. Only one polarity operation is shown for the sake of simplicity. Note that for the superbeam, the channels for the other polarity, which is present in the same beam as a contamination, are not explicitly shown.}
\end{table}

We use the following properties to parameterize the detector, and we will test the effects of each property separately:

{\bf Event mis-identification.} We assume that a certain fraction (``misID'') of the background event rate is mis-identified as signal. For the beta beam, this fraction applies to the neutral current and electron neutrino events.\footnote{The $\nu_e \rightarrow \tau_\tau$ oscillation channel has the same qualitative dependence on $\stheta$ as the $\nu_e \rightarrow \nu_\mu$ channel, which means that it adds to the $\stheta$ signal and hardly affects the measurement.} For the neutrino factory,  we take into account that the detector is non-magnetized. This means that $\nu_\mu$ and $\bar\nu_\mu$ events of the same beam polarity sum. The mis-identification fraction then applies to neutral current events, $\nu_e$, $\bar \nu_e$, $\nu_\tau$, and $\bar \nu_\tau$ (cascades). In addition, the muon decay mode of the taus ($\nu_\tau$ and $\bar \nu_\tau$) are included with a branching fraction of 17\% and the corresponding migration matrices, since these cannot be disentangled from muon tracks. For the superbeam, we include all neutral and charged current backgrounds. However, we distinguish two mis-identification fractions, one for muon tracks  (``misIDtracks'') and one for all the other flavors (``misID''). Here the logic is that  muon tracks may be not be as easily confused with other event topologies compared to cascades, \ie, misIDtracks $\le$ misID. While the disappearance channels are affected by misIDtracks, the appearance channels are affected by misID (apart from mis-identified muon tracks).  The intrinsic beam contaminations are fully taken into account.
The signal and backgrounds used for the simulation are summarized in \Tab~\ref{tab:backgrounds}.

{\bf Detection threshold.} We impose a sharp threshold on the detection, which corresponds to a sharp cutoff of the fiducial mass (energy bins below this cutoff are not taken into account). From the discussion in \Sec~\ref{sec:oscillation}, it is clear that such a threshold does not have a major effect below 2~GeV.

{\bf Energy resolution} We assume an energy resolution $\Delta E = \eta \cdot E$, where  $\eta$ is referred to as ``energy resolution''. We use a spectral analysis including the energy information, as well as we test a total event rate-based analysis.

{\bf Fiducial volume} Since the track and shower lengths increase with energy, the fiducial volume typically increases with energy. We test different fiducial masses, which we assume to be fixed over the whole energy range as a first approximation. We make a comparison  to preliminary \dc\ calculations for the fiducial mass as a function of energy in \App~\ref{app:fidvol}. Note that the main part of this study is independent of a specific detector, meaning it can also serve for the optimization of \dc\ phase~II.

\begin{table}[t]
\begin{center}
\begin{tabular}{llrrrrr}
\hline
Experiment & Setups & misID & misIDtracks & $E_{\mathrm{th}}$ & Energy res. & Fid. mass \\
\hline
Beta beam & \BBbm , \BBbms & $0.001$ & n/a & 2~GeV & $50\% \cdot E$ & 5~Mt \\
Neutrino factory & \NFbm , \NFbms & $0.001$ & n/a & 2~GeV & $10\% \cdot E$ & 5~Mt \\
Superbeam & \SBbms & $0.01$ & $0.01$ & 2~GeV & $20\% \cdot E$ & 5~Mt \\
\hline
\end{tabular}
\end{center}
\mycaption{\label{tab:standard}Benchmark values used for the \dc\ detector, together with the corresponding setup names. The values will be varied and studied individually in the following sections. }
\end{table}

In the next two sections we vary each of the above parameters to establish the minimal requirements for the detector. For each type of experiment, we choose a set of benchmark values, listed in \Tab~\ref{tab:standard}.
The logic of the setup names is the same as before: \BBbm\ and \NFbm\ correspond to \BBref\ and \NFref , respectively. The longer baseline detector is replaced by \dc .  \BBbms ,  \NFbms , and \SBrefs\ correspond to \BBrefs ,  \NFrefs , and  \SBrefs , respectively. The single baseline is replaced by a baseline to \dc . The benchmark setups (subscript ``bm'') are summarized in \Tab~\ref{tab:setups}.
 Note that some of the values (for example, the energy resolution) are not the same for all experiments. These choices reflect that the detector requirements for some experiments are more critical than for others. They are chosen to be in the regions of highest sensitivity for the detector performance.
The values for misID for beta beam and neutrino factory imply that muon tracks are identified with a contamination of less than 0.1\% from other flavors\footnote{As we will demonstrate later, a mis-identification of 1\%, which is perhaps more realistic, will also lead to reasonable sensitivities.}, whereas for the superbeam, such a low mis-identification rate is probably unrealistic (at least for electron neutrinos). The benchmark choice for the fiducial mass corresponds to the \dc\ fiducial mass for $\nu_\mu$ at about 4~GeV from preliminary simulations~\cite{Koskinen}. In either case, we will test the impact of these parameters in the following sections.

The systematic uncertainties are assumed to be the same for the beta beam and neutrino factory for consistency: 2.5\% signal normalization error and 5\% background normalization error. For the superbeam, we use 5\% and and 10\% for signal and background normalization, respectively, for the disappearance channels, as for the reference setups. For the appearance channels, we adopt a more conservative signal normalization error than for the reference setup, and we use the same numbers as for the beta beam and neutrino factory. Note that these effective systematic uncertainties arise from the beam flux normalization, the near-far extrapolation, and a fiducial mass error, just to name a few examples. Therefore, it is a consistent first estimate to use similar numbers for all experiments if the effective uncertainties are dominated by the detector.

\section{Requirements for two-baseline setups for small $\boldsymbol{\theta_{13}}$}
\label{sec:two}

Here we discuss the \dc\ requirements for two-baseline setups for a beta beam and neutrino factory. The second baseline is typically used for the mass hierarchy measurement and degeneracy resolution for small values of $\theta_{13}$. Compared to the reference setups in \Sec~\ref{sec:reference}, this second baseline is replaced by a core-crossing baseline to \dc . The performance is evaluated compared to the reference setups.

\subsection{Beta beam}
\label{sec:reqbb}

\begin{figure}[p]
 \centering
 \includegraphics[width=0.85\textwidth]{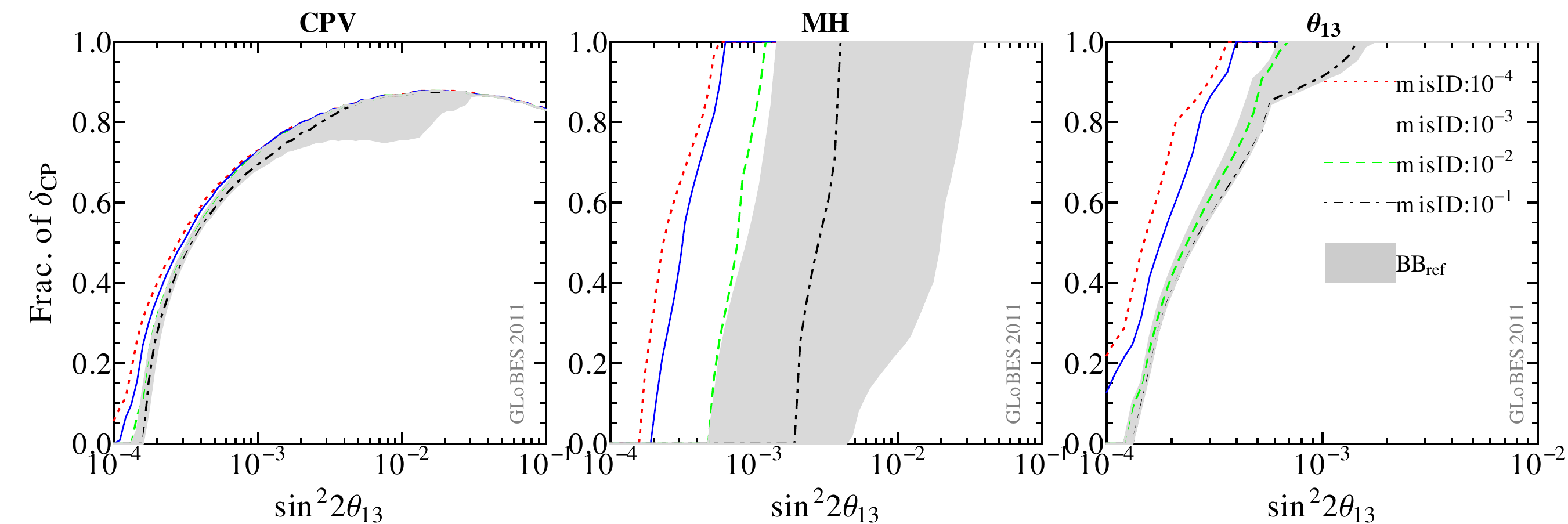}
 \includegraphics[width=0.85\textwidth]{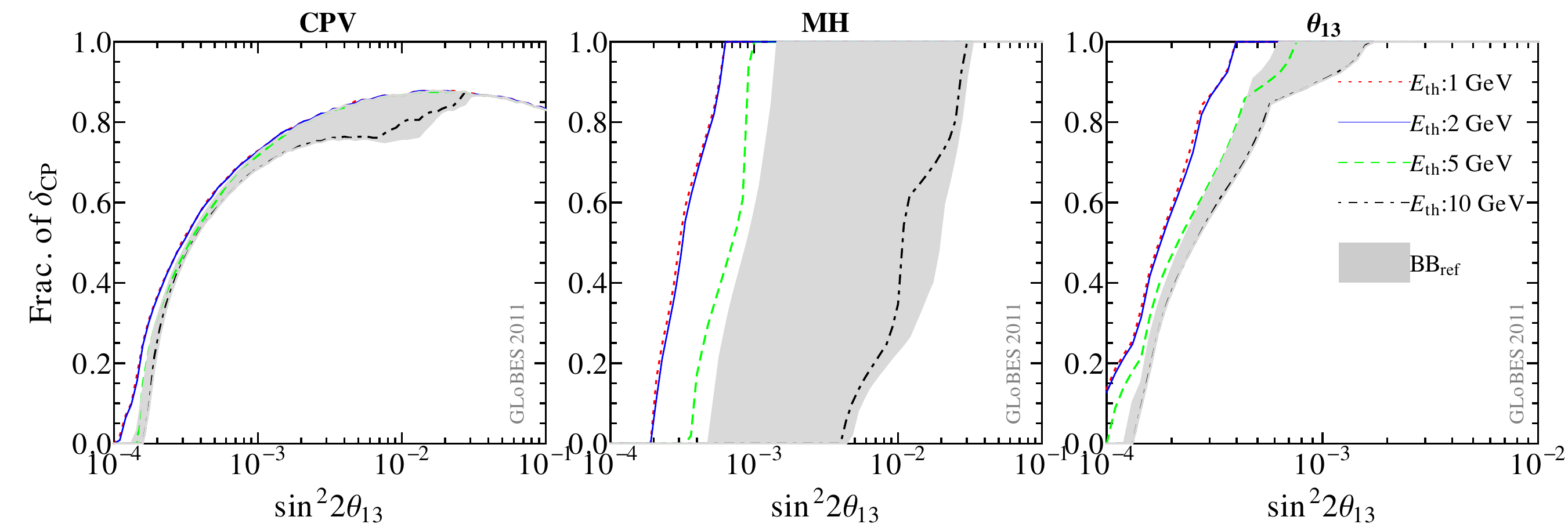}
 \includegraphics[width=0.85\textwidth]{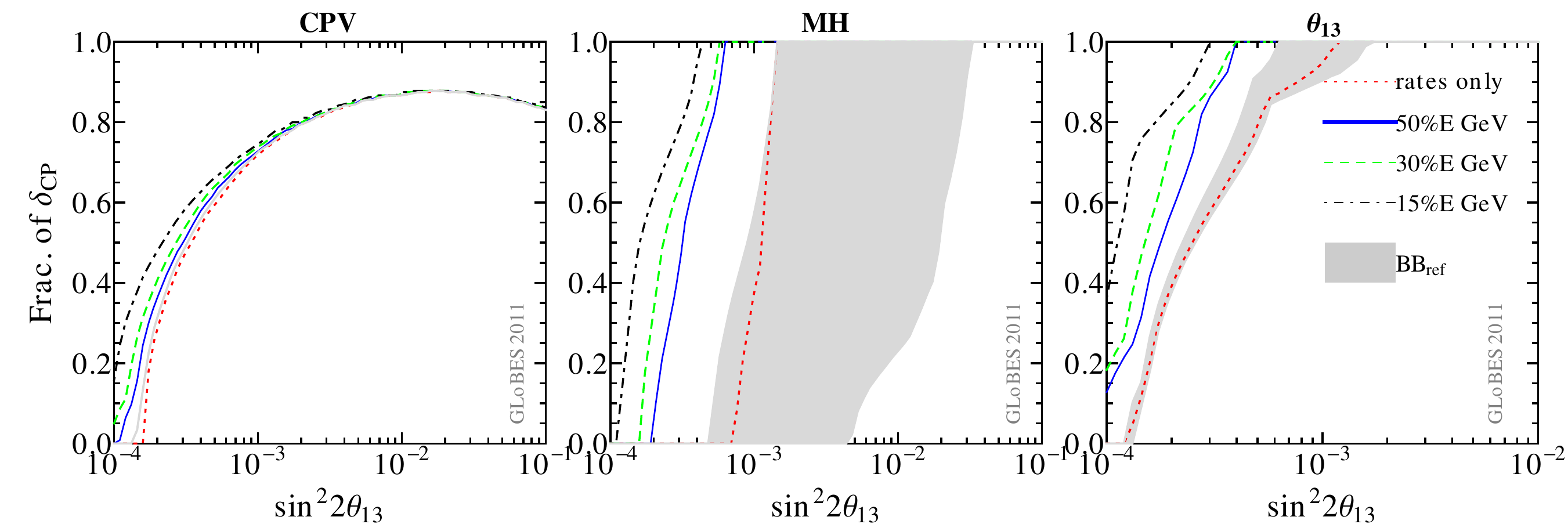}
 \includegraphics[width=0.85\textwidth]{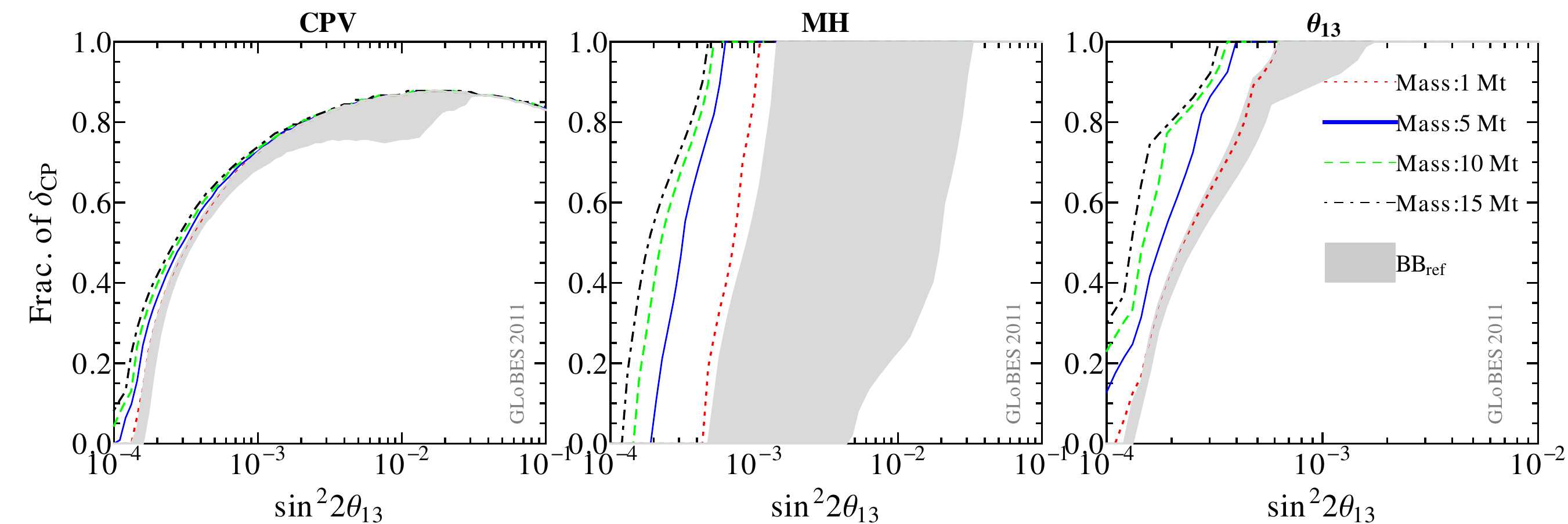}
 \mycaption{Discovery reach for CPV, MH and $\theta_{13}$ at $3\sigma$ for a two-baseline beta beam combination of WC detector and \dc\ detector. The solid curves correspond to the benchmark values from \Tab~\ref{tab:standard}, \ie, \BBbm . The detector parameters are individually varied in each of the rows and compared to the benchmark values. The shaded regions show the performance of the reference setup, with the right hand edges corresponding to the short baseline scenario, and the left hand edges corresponding to the baseline combination. Here we assume the normal hierarchy.}
 \label{fig:BB}
\end{figure}

The discovery reach of CPV, MH and $\theta_{13}$ at $3\sigma$ for a two-baseline beta beam combination of WC detector and \dc\ detector is shown in \figu{BB}. The solid curves correspond to the benchmark values from \Tab~\ref{tab:standard}. The detector parameters are individually varied in each of the rows compared to the benchmark values. The shaded regions show the performance of the reference setup, where the right edges of the shadings correspond to the short baseline only, and the left edges to the baseline combination. Since the long baseline of the reference setup was replaced by the \dc\ detector, approaching the right edges of the shadings means that the second baseline does not contribute. Exceeding the left edges of the shaded regions implies  the performance is better than that of the reference setup, which is regarded as an optimal setup for each class. A result well within the shaded area means the second baseline adds sensitivity, but less than the reference setup. 

Considering each of the detector parameters in turn, the mis-identification fraction (first row) should be smaller than about 0.01 to match the original reference setup (left edges of shadings), where the MH and $\theta_{13}$ reaches are most sensitive to this parameter. The energy threshold (second row) should be at most 5~GeV to match the reference setup, and a 2~GeV threshold leads to a significantly better performance than the reference. Beyond 2~GeV, a further reduction in energy threshold does not show additional improvement. This is an expected dependence from the neutrino appearance probability oscillation pattern illustrated in \figu{profile}.  The energy resolution (third row) should be at least about 50\% to outperform the reference setup. Note, however, that even with a total rate only analysis (``rates only'') the mass hierarchy performance of the reference setup is within reach. Better energy resolutions clearly improve the performance, far beyond the reference setup. The fiducial mass (lowest row) shows 1~Mt is sufficient to match the performance of the reference setup, and larger masses show clear improvement. All things considered, it may be useful to improve the mis-identification by appropriate cuts, at the price of a lower fiducial mass (efficiency).
From \App~\ref{app:fidvol}, where an energy-dependent fiducial mass is tested, it is clear that the values for threshold and fiducial mass used in \Tab~\ref{tab:standard} match such an energy-dependent fiducial mass rather well, and are, in this case, possibly conservative.

In summary, the detector requirements for \dc\ as a beta beam detector at a very long baseline are modest. The primary issue  is a relatively clean flavor-identification of muon neutrinos, which may be achieved by selecting events with some loss of efficiency. On the other hand, energy resolution is not so important.

\subsection{Neutrino factory}
\label{sec:reqnf}

\begin{figure}[t]
 \centering
 \includegraphics[width=0.9\textwidth]{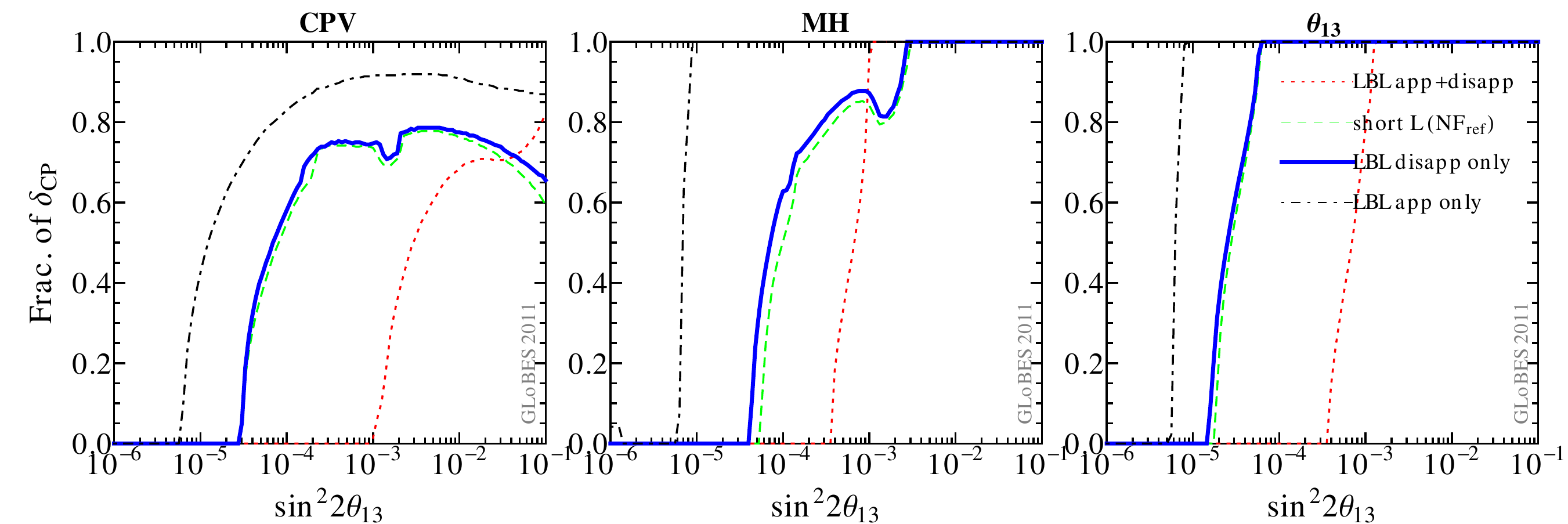}
 \mycaption{Discovery reach for CPV, MH and $\theta_{13}$ at $3\sigma$ for different channel configurations of a single-baseline neutrino factory, where a (hypothetical) magnetized detector at the very long \dc\ baseline is assumed and compared to the short baseline of \NFref .  Values from \Tab~\ref{tab:standard} are used.
}
 \label{fig:NFchannels}
\end{figure}

A peculiarity for the neutrino factory is the assumption of a non-magnetized detector. This means that, for instance, the muon neutrino and antineutrino event rates from Eqs.~(\ref{equ:numu}) and~(\ref{equ:numubar}) must be added, and appearance and disappearance channels can only be discriminated by energy resolution, as illustrated previously in  \figu{probcomp}. However, note that although not as efficient, the disappearance channel also has some sensitivity to $\stheta$ and $\deltacp$, as discussed in \Sec~\ref{sec:oscillation}. We show this effect quantitatively in \figu{NFchannels} for a single-baseline neutrino factory, where the  MIND detector performance at the short baseline is shown for comparison. In this figure, we hypothetically assume that the appearance and disappearance channels could be separated, \ie, the detector would be magnetized. Obviously,  the very long baseline has an excellent reach in the appearance channels because of the enormous statistics, and even the disappearance channels can measure all of the performance indicators better than the shorter baseline MIND setup. Of course, since the relevant terms are sub-leading in the disappearance channels, the performance is significantly worse than for the appearance channels. However, if the appearance and disappearance information is added, the information on CPV, MH and $\theta_{13}$ is significantly reduced compared to either of the channels alone (unless $\theta_{13}$ is very large, where the signal in the disappearance channel dominates). This is the situation in a non-magnetized detector, such as \dc . 

\begin{figure}[mp]
 \centering
 \includegraphics[width=0.9\textwidth]{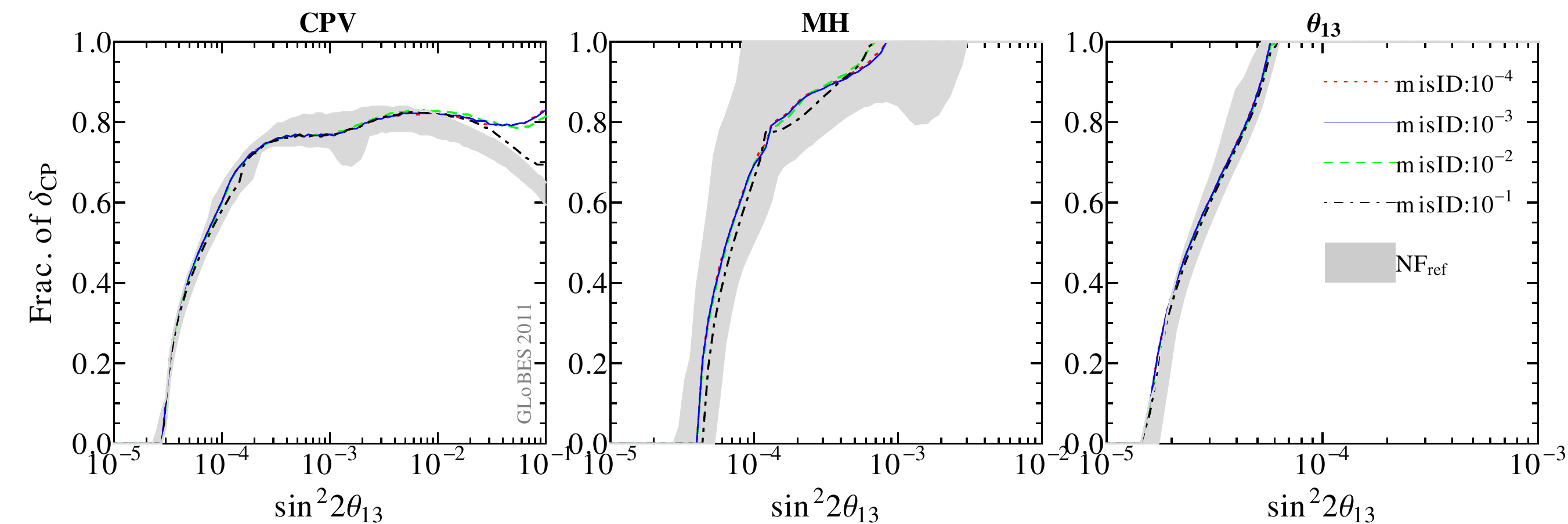}
 \includegraphics[width=0.9\textwidth]{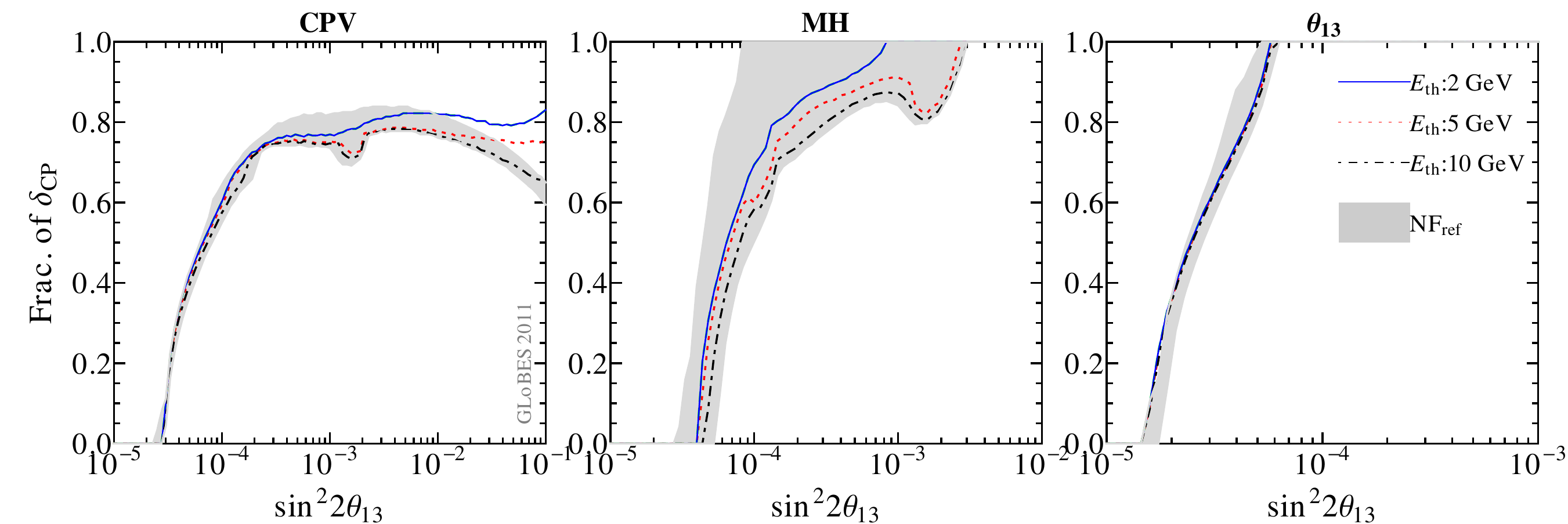}
 \includegraphics[width=0.9\textwidth]{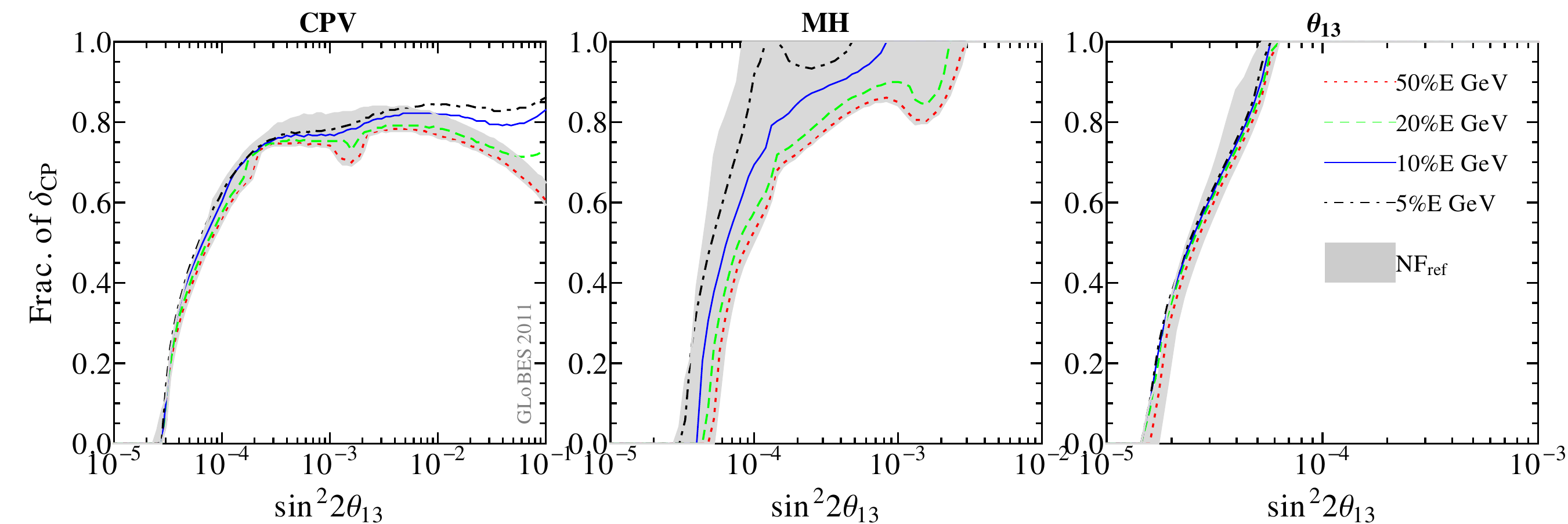}
 \includegraphics[width=0.9\textwidth]{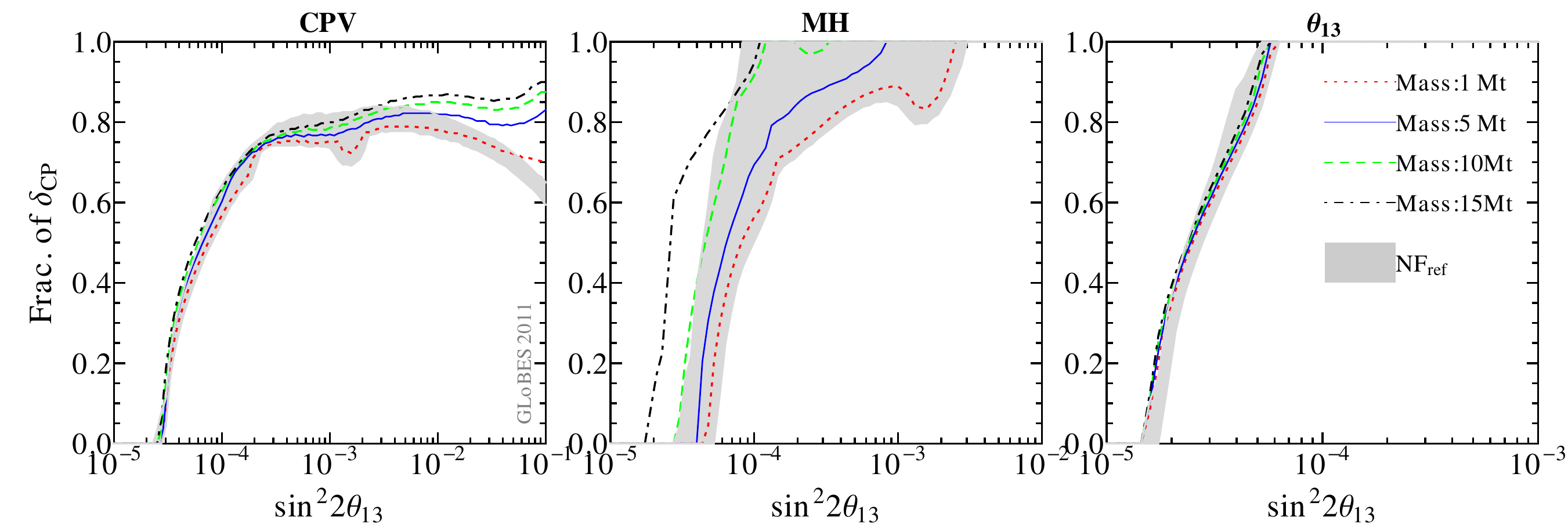}
 \mycaption{Discovery reach for CPV, MH and $\theta_{13}$ at $3\sigma$ for a two-baseline neutrino factory combination of MIND detector and \dc\ detector. The solid curves correspond to the benchmark values from \Tab~\ref{tab:standard}, \ie, \NFbm . The detector parameters are individually varied in each of the rows and compared to the benchmark values. The shaded regions show the performance of the reference setup, with the right hand edges corresponding to the short baseline scenario, and the left hand edges corresponding to the baseline combination. Here we assume the normal hierarchy.
}
 \label{fig:NF}
\end{figure}

Similar to the beta beam discussion, we show in \figu{NF} the requirements for the long baseline non-magnetized \dc\ detector as a second baseline detector.
As discussed above, the most critical requirement for the (non-magnetized) neutrino factory detector is energy resolution, where 50\%, as used for the beta beam benchmark, is clearly shown to be insufficient. We therefore apply a smaller value (10\%) for the benchmark setup in \Tab~\ref{tab:standard}. On the other hand, at least in principle, the reference setup can be roughly matched for (a hypothetical) 5\% energy resolution for small $\theta_{13}$, whereas already for 20\% energy resolution a better performance is obtained for CPV for large $\stheta$. The dependence on the threshold (second row) is qualitatively similar to the beta beam, since it depends on the coverage of the matter effect peaks. In particular, for small $\theta_{13}$, another critical factor is the fiducial mass (lower row), where at least 15-20~Mt is needed for a sufficient CPV performance. The reason is that for small $\theta_{13}$ the appearance signal can only be disentangled from the large disappearance background with sufficient statistics. On the other hand, the $\stheta$ terms in the disappearance channel also be contaminated by a large background from the leading oscillation terms. The mis-identification fraction (first row) is of minor importance, as long as it is, at most, 0.01, because the intrinsic background from the disappearance channel is dominant.
If an energy-dependent fiducial mass is used, see \App~\ref{app:fidvol}, our estimates in \Tab~\ref{tab:standard} are conservative. Compared to the beta beam, the neutrino factory spectrum peaks at higher energies, where there actual fiducial mass is larger than 5~Mt. Note that we have also included the intrinsic $\nu_\tau$ contamination in this analysis. This may be potentially harmful since the muon to tau neutrino oscillation sample, which comes from a leading oscillation channel driven by the large $\theta_{23}$, may be reconstructed at energies where the sub-leading  appearance channel, driven by the small $\theta_{13}$, dominates. 

In summary we find that the critical requirements for the non-magnetized \dc\ as a neutrino factory detector are energy resolution and statistics. Due to the higher beam energy of a high energy neutrino factory, large fiducial masses may be easily reached. However, we note the required energy resolution at a 10\% level may be unrealistic for muon tracks in such a detector.

\section{Single-baseline setups for large $\boldsymbol{\theta_{13}}$}
\label{sec:one}

In this section we discuss single-baseline setups for large $\theta_{13}$. We start with the superbeam, similar to the previous section, and briefly discuss the optimization of neutrino factory and beta beam for large $\theta_{13}$. Finally, we compare the precision measurements of the single-baseline \dc\  setups to the corresponding reference setups. 

\subsection{Requirements for the superbeam}
\label{sec:reqsb}

\begin{figure}[mp]
 \centering
 \includegraphics[width=0.9\textwidth]{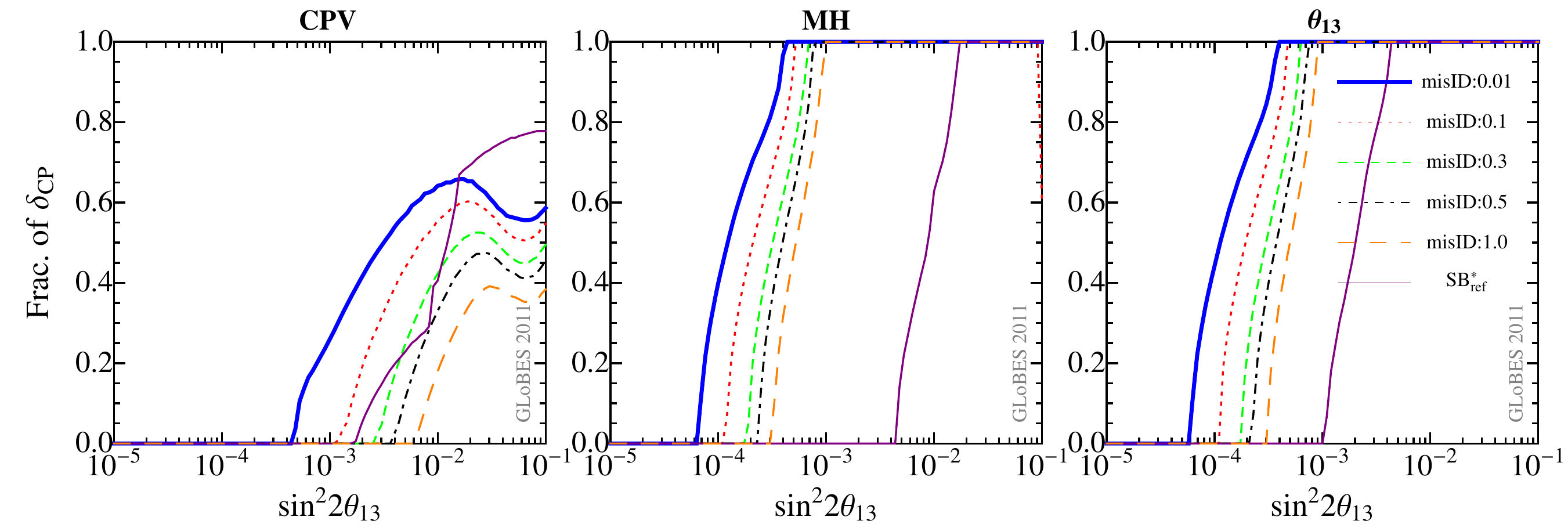}
 \includegraphics[width=0.9\textwidth]{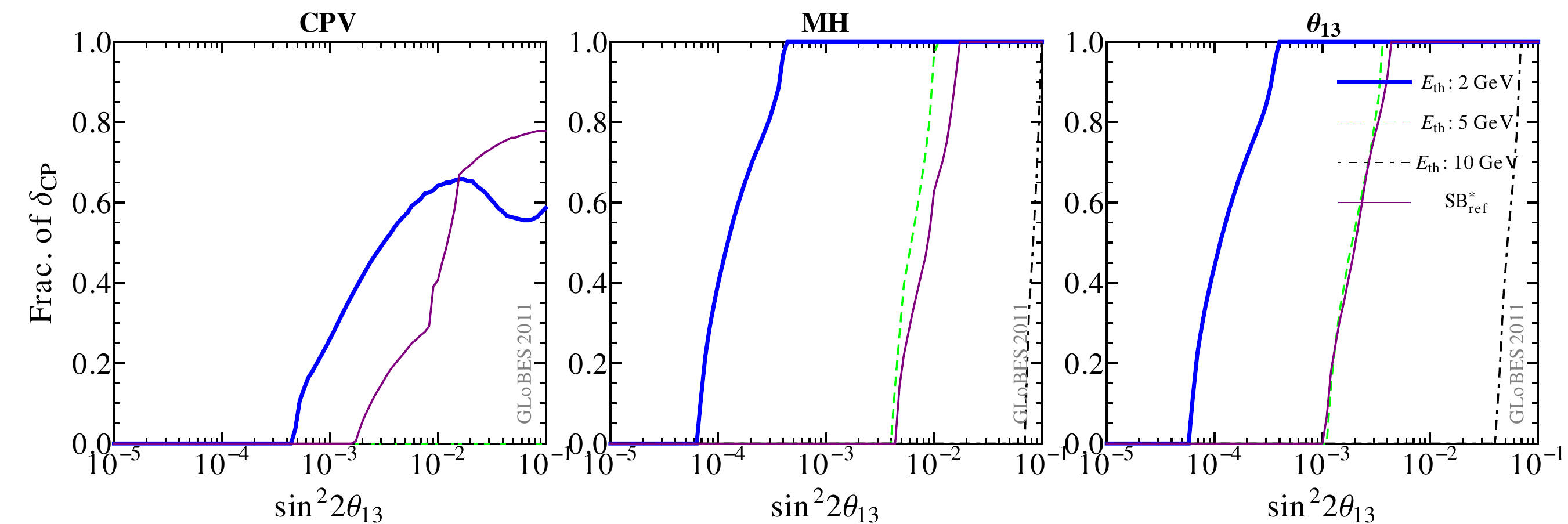}
 \includegraphics[width=0.9\textwidth]{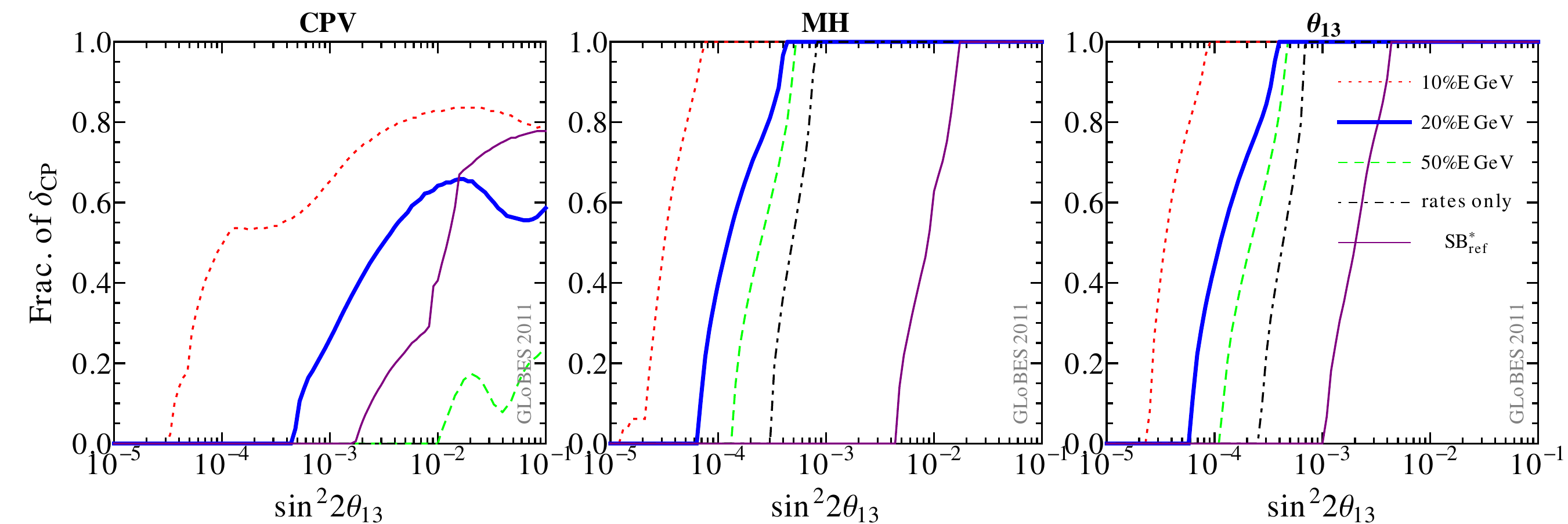}
 \includegraphics[width=0.9\textwidth]{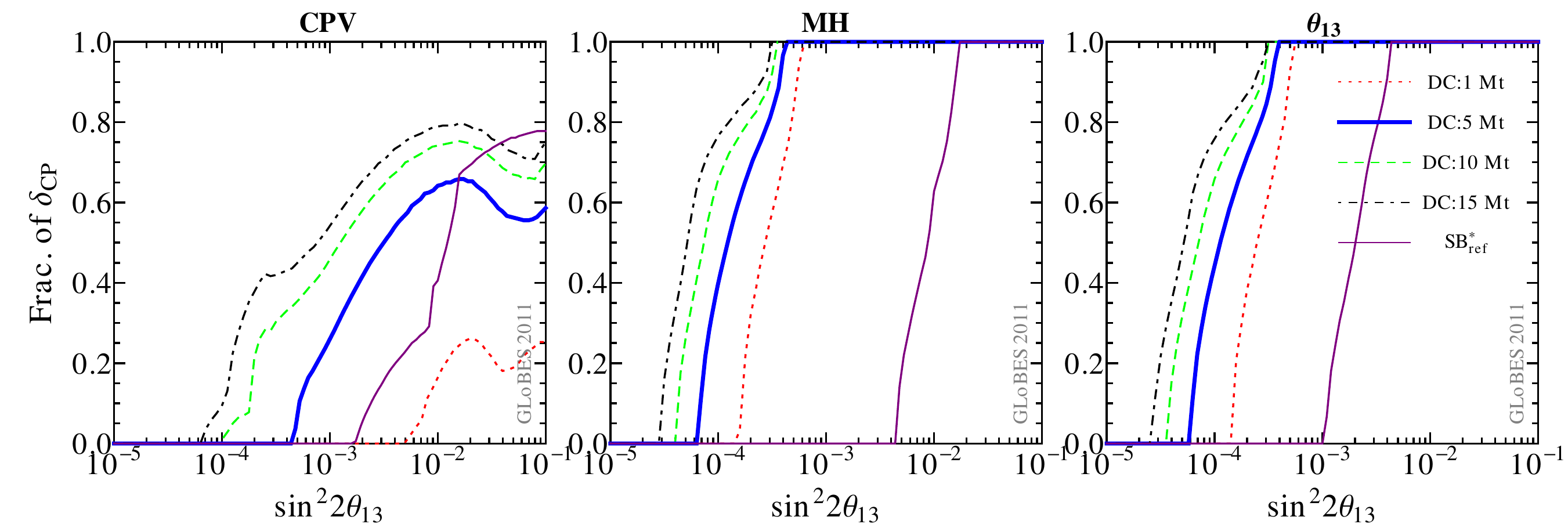}
 \mycaption{Discovery reach for CPV, MH and $\theta_{13}$ at $3\sigma$ for a single-baseline superbeam with core-crossing baseline to the \dc\ detector. The thick solid curves correspond to the benchmark values from \Tab~\ref{tab:standard}, \ie, \SBbms . The detector parameters are individually varied in the different rows and compared to the benchmark values. The thin solid curves show the performance of the reference setup. Here we assume the normal hierarchy.}
 \label{fig:LBNE}
\end{figure}

Similar to the previous section, we illustrate the requirements for \dc\ as a superbeam detector in \figu{LBNE}. As far as the mis-identification is concerned, note that we distinguish the identification of muon tracks, as needed for the disappearance channels, and the identification of electron neutrinos (cascades), as needed for the appearance channels. Since the appearance channels are critical for the measurements in \figu{LBNE}, it can be shown that the challenge is the identification of the electron neutrinos. We show in \figu{LBNE}, upper row, the corresponding mis-identification fractions. From our tests we find the identification of the muon tracks is not critical as long as the corresponding mis-identification fraction misIDtracks $\lesssim$ misID, while misIDtracks $\ll$ misID does not significantly improve the results. We therefore choose 0.01 for misIDtracks in \figu{LBNE}, showing larger values for misID. Note that the discussed mis-identification fractions are substantially larger than the ones for a beta beam and neutrino factory, because it could be difficult to assign a specific flavor to the cascades. The case for which the flavor of a cascade cannot be assigned corresponds to misID=1.0. From \figu{LBNE} it can be clearly seen that a small mis-identification fraction, less than 30\%, is necessary to match the performance of the reference beta beam for small $\theta_{13}$ for CPV. Interestingly, for MH and $\theta_{13}$ the performance of the reference superbeam is always exceeded even if the cascades cannot be identified. However, the reference CPV performance for large $\theta_{13}$, \ie, the case for which one would probably build a superbeam, is significantly difficult to achieve because of the smaller CPV effects at the long baseline. The reason for the ``inverted'' behavior compared to the reference is that the very long baseline is not very sensitive to $\deltacp$ and the leading term proportional to $\stheta$ dominates. Note that the statistics from the large fiducial mass helps for small $\theta_{13}$. 

As shown in \figu{LBNE}, the dependencies on threshold, energy resolution, and fiducial masses are all critical for the CPV measurement. For MH and $\theta_{13}$, typically moderate requirements, such as a threshold of 5~GeV and a rates only measurement, are sufficient to match the performance of the reference setup. Note, however, that for large $\stheta \gtrsim 0.01$, the reference superbeam could also measure these parameters for any value of $\deltacp$. In order to reach a competitive CPV performance, a relatively large fiducial mass is needed for large $\theta_{13}$. However, as seen in \App~\ref{app:fidvol}, the standard curves already reasonably represent the  energy-dependent fiducial mass due to the moderately high beam energy. 

In summary, the requirements for \dc\ as a single-baseline superbeam detector are significant. In particular, for the CPV measurement, good flavor identification of electron neutrinos with less than 1\% mis-identified events, a large fiducial mass, and a good energy resolution are required. Therefore, unless a much larger detector is planned for a second phase of \dc , the experiment may be challenging. The parameter choices in \Tab~\ref{tab:standard} reflect that result and we note that such a superbeam experiment could serve as a second baseline for the mass hierarchy measurement only.

\subsection{Minimal beta beam and neutrino factory setups}
\label{sec:minimal}

While the superbeam is designed as  an instrument for large $\theta_{13}$, a neutrino factory and a beta beam are typically optimized in a different way. The idea is to propose a ``minimal'' version of these experiments which can perform each of the measurements. In particular, only a single-baseline may be used, and lower muon energies (neutrino factory) or $\gamma$ (beta beam) may be allowed. As with the single-baseline superbeam, the problem is then,  of course, that the very long baseline is not optimal for CPV measurements.

We have studied the $E_\mu$ and $\gamma$ dependence for the neutrino factory (\NFbms) and the beta beam (\BBbms), respectively. For the beta beam, \br and \li have to be used since the relevant oscillation features at the very long baseline otherwise can not be studied fully. We have re-scaled the boost factors of both isotopes. For smaller boost factors the CPV discovery reach is severely affected, whereas larger boost factors may not be possible even at an upgraded CERN-SPS. Therefore, we use the earlier proposed $\gamma = 656$ and $\gamma = 390$ for $^8$B and 
$^8$Li, respectively.

\begin{figure}[!t]
 \centering
 \includegraphics[width=0.5\textwidth]{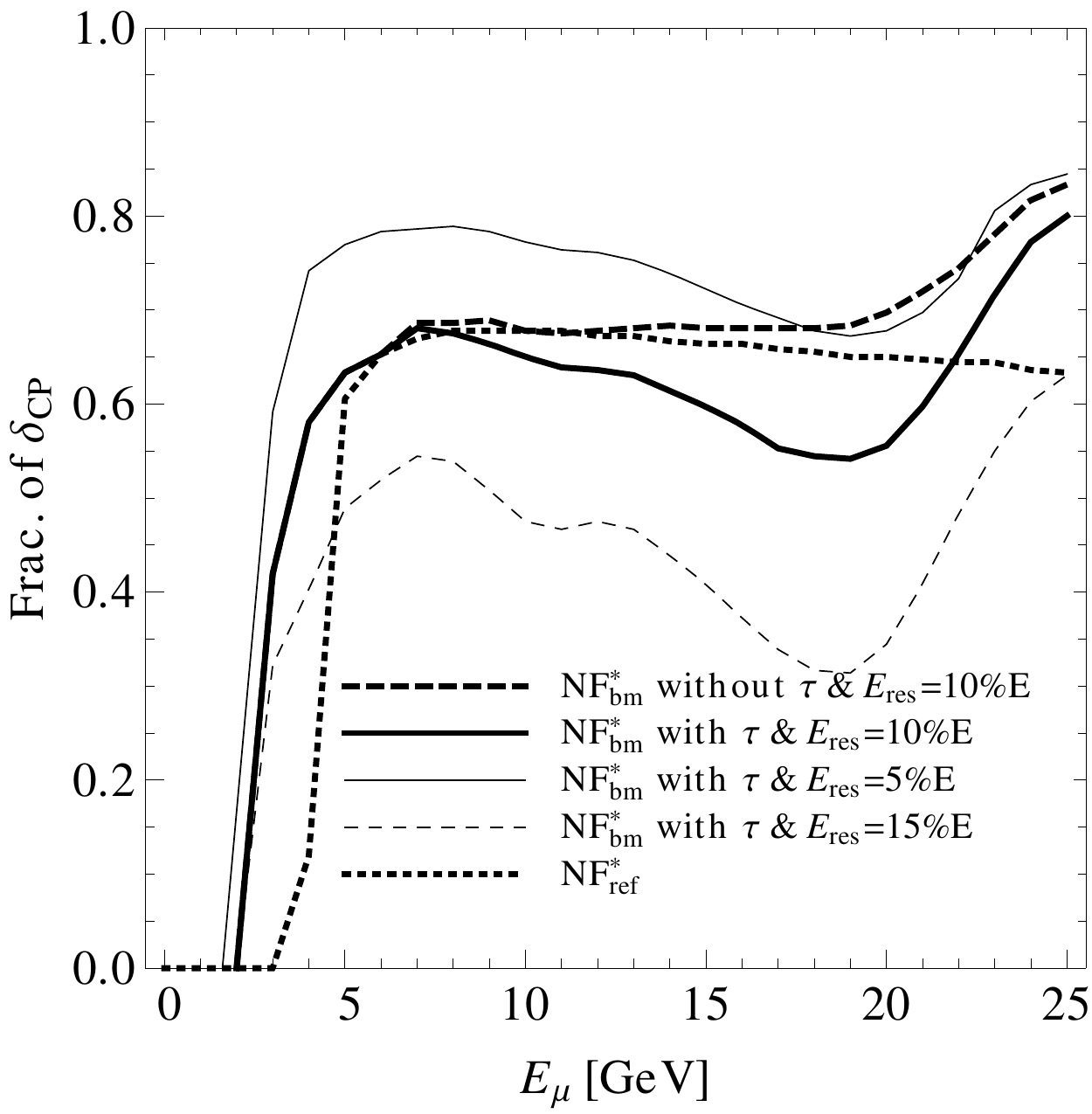}
 \mycaption{Fraction of $\deltacp$ for which CPV will we observed ($3 \sigma$) as a function of $E_\mu$ for $\sin^22\theta_{13}=10^{-1}$. Here the energy resolution is varied, and the effect of the $\nu_\tau$ backgrounds is shown. See \Tab~\ref{tab:standard} for the rest of the benchmark values used. The label $\mathrm{NF}_{\textrm{ref}}^*$ refers to the low energy ($E_\mu=10 \, \mathrm{GeV})$ reference version of a neutrino factory with MIND at 2200~km. Here we assume the normal hierarchy.}
 \label{fig:NF-Echeck}
\end{figure}

For a neutrino factory, the mass hierarchy and $\theta_{13}$ can typically be measured at the very long baseline for large $\theta_{13}$ and the critical factor is the CPV discovery reach. In \figu{NF-Echeck} the dependence of the fraction of $\deltacp$ for the CPV discovery as a function of $E_\mu$ for a large value of $\stheta=0.1$ is shown, which is close to the current bound. For the reference setup (MIND@2200~km) it can be clearly seen that $E_\mu$ around 10~GeV is optimal, for which CPV can be measured for at least 65\% of all $\deltacp$. It is instructive to study the curve ``\dc\ without $\tau$'' with 10\% energy resolution and without the intrinsic $\tau$ contamination ($\nu_\tau$ or $\bar\nu_\tau$ producing tau leptons, which decay into muons). This curve monotonously increases with $E_\mu$ and plateaus between about 7 and 18~GeV. This behavior can be understood in terms of \figu{probcomp}: the $P_{e \mu}$-dominated range near 7.5~GeV is covered by the spectrum, where the discovery reach will be saturated.  It is interesting to note the behavior  if the $\tau$ contamination is included. The curve is similar to the one without $\tau$ contamination until about 6~GeV, since the $\tau$ production threshold is not significantly exceeded, and the performance then decreases. This is due to  17\% of the $\nu_\tau$ from the leading $\nu_\mu \rightarrow \nu_\tau$ channel being mis-identified  as $\bar \nu_\mu$ adding to the $\bar \nu_e \rightarrow \bar\nu_\mu$ appearance channel in the non-magnetized detector. Note that these events are  reconstructed at lower energies, which means that the ranges in \figu{probcomp}, where the appearance information dominates, are affected. Only if $E_\mu$ approaches 25~GeV, the performance does increase again. Therefore, for the LENF to \dc , we choose $E_\mu = 5 \, \mathrm{GeV}$, where the $\tau$ contamination issue can be completely avoided and the comparison with the reference setup shows a similar performance. As it can be seen from the curves with different energy resolutions, this choice is at the lower end of the allowed $E_\mu$ range, but closer to the original idea of the LENF. From the 15\% energy resolution curve it can be also seen that the $\tau$ contamination must be avoided in that case; here the power of the energy resolution to discriminate different channels is severely limited by the different reconstructed energies of the events from $\tau$ decays. 

\subsection{Precision measurements with single-baseline setups}

\begin{figure}[!t]
 \centering
 \includegraphics[width=0.9\textwidth]{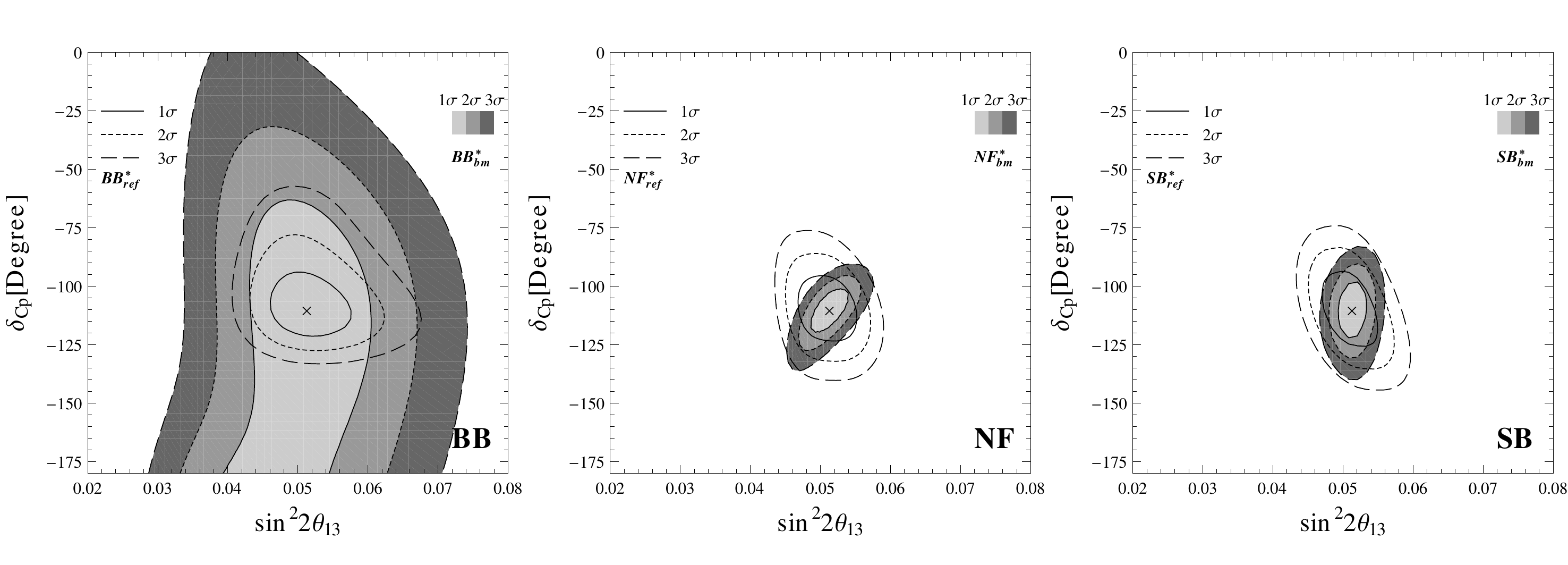}
 \mycaption{The precision measurements of CP phase $\delta_{\mathrm{CP}}$ and $\stheta$ for three single-baseline neutrino experiments:  Beta Beam (BB), Neutrino Factory (NF), and SuperBeam (SB). The contours represent the  $1\sigma$, $2\sigma$ and $3\sigma$ confidence levels (2 d.o.f.). Filled contours represent the \dc\ benchmark setups and unfilled contours the reference setups. The crosses mark the best fit value of $\sin^22\theta_{13}$ and $\delta_{\mathrm{CP}}$.  Here we assume the normal (true) hierarchy. The inverted (fit) hierarchy solution can be ruled out by the experiments.}
 \label{fig:precision}
\end{figure}

After the recent release of long-baseline electron neutrino appearance searches at the T2K and MINOS experiments, a global analysis hints at a $1\sigma$ level to the Dirac CP violating phase~\cite{Schwetz:2011zk}. We also consider the best fit values of $\sin^2\theta_{13}=0.013$ and $\delta_{\mathrm{CP}}=-0.61\pi$ for the normal hierarchy and study precision measurements at three neutrino experiments with different single-baseline beams in \figu{precision}. The single-baseline reference setups are defined in \Sec~\ref{sec:reference}, whereas the \dc\ versions are described earlier in this section. Note that there is only a fit for one hierarchy in the figure, because the inverted hierarchy solution can be excluded. 

From \figu{precision} it can be seen that the \dc\ setup can not only compete with neutrino factory and superbeam reference, but,  in fact, has performance which is somewhat better. For the neutrino factory, this could be see already in \figu{NF}, where the long baseline helped for $\deltacp$ if $\stheta$ was large. For the superbeam, the \dc\ setup in \figu{precision} (right panel) appears to be better, which is opposite the behavior seen in \figu{LBNE}, where. for large $\theta_{13}$, the reference setup is better. The reason for this is the different dependence on $\deltacp$. The CPV performance roughly corresponds to the precision at $\deltacp=0$ or $\pi$, but the precision shown in \figu{precision} is depicted for $\delta_{\mathrm{CP}}=-0.61\pi$. We confirm that for $\deltacp=0$ the performance of \figu{LBNE} is recovered, meaning the direct comparison of the two setups strongly depends on $\deltacp$. For a relatively narrow beam spectrum in particular, depending on the baseline, the $\sin \deltacp$- or $\cos \deltacp$-term can be emphasized, see \equ{papp}, leading to  a strong dependence on $\deltacp$.

For the beta beam, it was already obvious from \figu{BB} that the CPV does not significantly benefit from the very long baseline. This explains why the CPV is significantly worse than for the shorter baseline reference in \figu{precision}. Compared to neutrino factory and superbeam, the main issue here is the relatively asymmetric operation with $\nu_e$ and $\bar \nu_e$ with very different peak energies, whereas the muon energies for $\nu_e$ and $\bar \nu_e$ at the neutrino factory are exactly alike. This is particularly true since the $\nu_e$ spectrum peaks at relatively low energies and the  disappearance channel is missing.

In summary, for large $\theta_{13}$,  \dc\ may be an interesting option to receive a beam sent from the LENF or a seperbeam. We note, however, that the values required for the detector performance are relatively challenging, see \Tab~\ref{tab:standard}. For instance, a very good energy resolution is required for both setups. On the other hand, the beta beam, for which the detector performance is of secondary importance, suffers from constraints on the boost factors from the CERN-SPS. A balanced green-field scenario (\ie, a setup with a new accelerator) may, however, have a better performance.

\section{Measurement of the Earth's core density}
\label{sec:earth}

\begin{figure}[!t]
 \centering
 \includegraphics[width=0.9\textwidth]{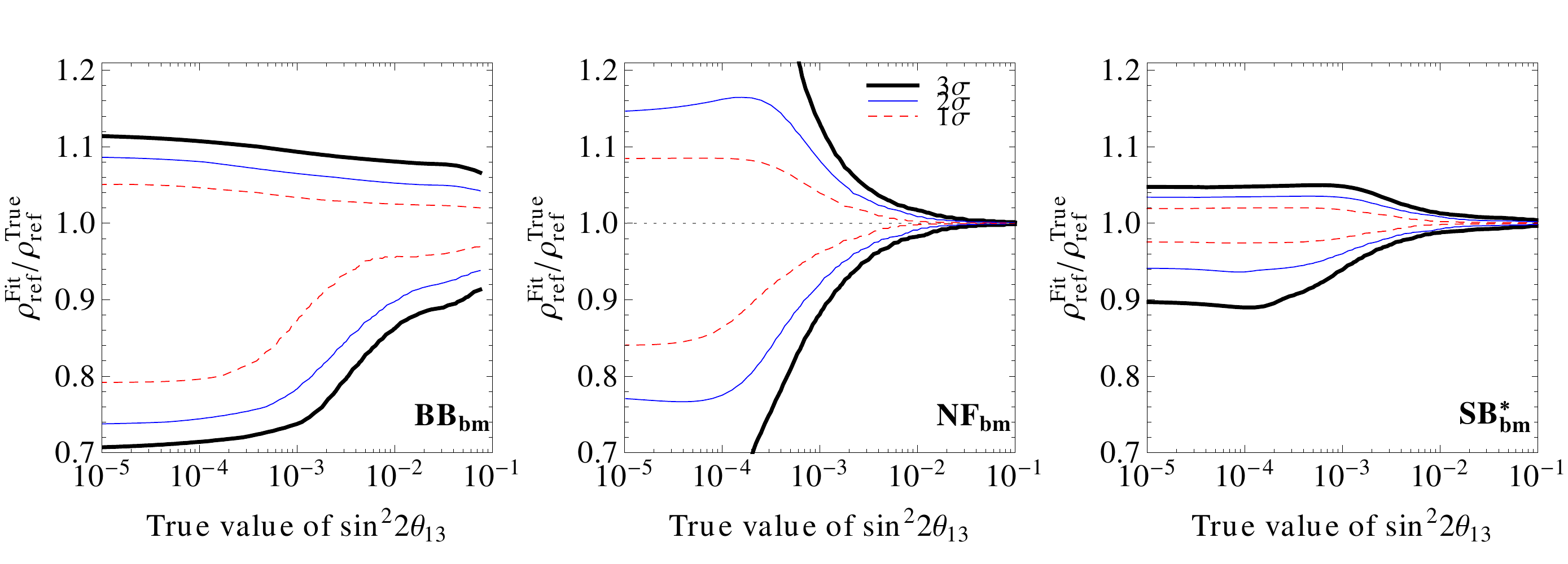}
 \mycaption{Relative precision of the Earth's core density measurement as a function of the true value of $\sin^22\theta_{13}$ for three neutrino experiments with different beams at the $1\sigma$, $2\sigma$ and $3\sigma$ confidence levels. Here we assume the normal hierarchy with $\deltacp=0$. 
}
 \label{fig:matter}
\end{figure}

A feature of the \dc\ setups discussed in this work is that they cross the Earth's outer core. It may be possible to use this to determine the core density by the MSW effect~\cite{Winter:2005we}. Here we discuss the quantitative performance for such a measurement of the two-baseline (beta beam, neutrino factory) or single-baseline (superbeam) benchmark setups, discussed earlier in this work. The density profile of the Earth for the long baseline is approximated by the three-step profile in the left panel of \figu{profile}. For the sake of simplicity, we assume that the Earth's mantle density is relatively well known from seismic wave geophysics, and that the core density, \ie\ the higher density in \figu{profile}, is to be measured.\footnote{See \Ref~\cite{Gandhi:2006gu} for the test of more complicated hypotheses.} In this measurement, the oscillation parameters are marginalized over, which means that the uncertainties on the oscillation parameters, as obtained from the experiment itself, are included.

We show in \figu{matter} the relative matter density precision for the (outer) core density measurement as a function of the (true) $\stheta$, assuming that the chemical composition is known. For large $\stheta$, the information in the appearance channel is typically clean enough  for a relatively precise measurement (the backgrounds from the second to fourth terms in \equ{papp} are small).
 For intermediate  values of $\stheta$, correlations and degeneracies prohibit the measurement of the density. For small values of $\stheta$, the measurement is dominated by the solar term; \cf, last term in \equ{papp}. 

For the neutrino factory, the core density can be determined to an accuracy of up to $0.5\%$ at $1\sigma$ for  $\stheta \gtrsim 10^{-2}$.  For small values of $\stheta$, however, the solar term cannot be easily extracted because of the disappearance channel contamination in the non-magnetized detector. For the superbeam, the performance is similar to that the neutrino factory, and to predictions  from past studies~\cite{Gandhi:2006gu,Minakata:2006am}.  Precise numbers at different  confidence levels are found in \Tab~\ref{tab:matter}, assuming that $\sin^22\theta_{13}=10^{-2}$.
As a major difference of superbeam and beta beam compared to the neutrino factory, the matter density can be also measured relatively well for small values of $\stheta$ because the appearance and disappearance channels can be separated.  However, for large $\stheta$, the statistics of the beta beam are not comparable to that of neutrino factory and superbeam, for the reasons mentioned in the previous subsection.

\begin{table}[t]
\centering
\begin{tabular}{|c|ccc|}
\hline
 &&\text{Measurement of the core density $\varepsilon$}&\\
 &\text{1$\sigma $} & \text{2$\sigma $} & \text{3$\sigma $} \\
\hline
\multirow{2}{*}{$\text{BB}_{\text{bm}}$}
 &+0.0263 & +0.0532 & +0.0807 \\
 &-0.0461 & -0.1030 & -0.1380 \\
\hline
\multirow{2}{*}{$\text{NF}_{\text{bm}}$} 
&+0.0058 & +0.0117 & +0.0174 \\
 &-0.0059 & -0.0120 & -0.0181 \\
\hline
\multirow{2}{*}{$\text{SB}_{\text{bm}}^*$} 
 &+0.0046 & +0.0091 & +0.0137 \\
 &-0.0045 & -0.0089 & -0.0132 \\
\hline
\end{tabular}
\mycaption{\label{tab:matter}The relative precision of core density measurement  $\varepsilon\equiv(\rho_{\mathrm{ref}}^{\mathrm{Fit}}-\rho_{\mathrm{ref}}^{\mathrm{True}})/\rho_{\mathrm{ref}}^{\mathrm{True}}$ at different confidence levels assuming $\sin^22\theta_{13}=10^{-2}$.}
\end{table}

\section{Summary and conclusions}

A future upgrade of the IceCube/DeepCore detector at the South Pole, known as \dc\ (``Precision IceCube Next Generation Upgrade''), would have a lower threshold via denser photomultiplier spacing, making it, in principle, an interesting candidate for a neutrino beam. While the baseline from any of the major accelerator laboratories in the Northern hemisphere is not optimal for CP violation studies, the huge increase of the detector fiducial mass, compared to conventional detectors, may compensate. We have tested the requirements for such a detector, which is currently in the design phase. As a starting point, we have used beta beam, neutrino factory, and superbeam setups which are accepted to be optimized for either small (undiscovered) or large (discovered) $\stheta$. For the small $\theta_{13}$ setups, we have replaced the second ``magic'' baseline by a baseline to \dc .  For large $\theta_{13}$, confirming the recent results from T2K and Double Chooz, we have discussed ``minimal'' single-baseline versions. In all cases, we have compared the performance quantitatively to the known optimized reference setups in order to determine the minimal requirements for the \dc\ detector. As main performance indicators, we have used the conventional discovery reaches for CP violation, mass hierarchy, and $\theta_{13}$. For large $\theta_{13}$, we have also shown the precision for the $\theta_{13}$ and $\deltacp$ measurements.

From the point of view of neutrino oscillations, the baseline to \dc\ is interesting since it crosses the Earth's (outer) core. This implies the appearance channel is parametrically enhanced, see \figu{profile}, with resonance energies corresponding to the Earth's mantle and core densities. This has immediate consequences for the detection threshold of the \dc\ detector and the required beam energies: the core resonance requires a measurement between 2~and 5~GeV, and the mantle resonance between 5~and 10~GeV. Thresholds lower than 2~GeV or  significantly above 10~GeV minimally impact the results. The chosen beams in this study take this energy range into account, and it is coincidentally a range where the \dc\ fiducial mass is large.

For small values of $\theta_{13}$, a scenario where the recent T2K and Double Chooz results are not confirmed and \dc\ has replaced the magic baseline detector, we have identified that the flavor-clean beta beam imposes the least requirements to the detector. Most important is a relatively  clean flavor-identification of muon neutrinos. Using muon tracks, this may be achieved by applying strict cuts at the expense of fiducial mass. Energy resolution is of secondary importance. The neutrino factory, on the other hand, requires channel identification by energy resolution if the detector is non-magnetized. The required energy resolution of about $10 \% \, E$ may prove challenging for short muon tracks in the \dc\ detector. 

For large $\theta_{13}$ ($\theta_{13}$ discovered), \dc\ has been studied as a single-baseline standalone detector of a long-baseline experiment. For a superbeam, using the LBNE beam as a guide, the requirements are high for CPV measurements: 
flavor identification of electron neutrinos with less than 1\% mis-identified events from other flavors, a large fiducial mass, and a quality energy resolution are needed. Unless a much larger detector is planned for a phase~II of \dc\ or the detector is used for the mass hierarchy measurement only, the experiment may be challenging. For the beta beam, the CPV has found to be strongly constrained by the limited boost factors in an upgraded CERN-SPS. However, a green-field scenario not restricted to this particular accelerator may have better performance. For the neutrino factory, energy resolution is a major challenge, similar to the small $\stheta$ case. However, small muon energies as low as $E_\mu \simeq 5 \, \mathrm{GeV}$ are preferable, where the intrinsic $\tau$ contamination (muon tracks from tau decays) can be avoided. This problem has turned out to be especially severe in a non-magnetized detector, where the secondary muons (from the tau decays into muons) from the leading channel $\nu_\mu \rightarrow \nu_\tau$ may be reconstructed in energy windows where the appearance channel is to be isolated by energy resolution. Such a low muon energy may be especially interesting in connection with recent discussions on a low luminosity and low energy initial stage neutrino factory.

We have also studied the potential measurement of the Earth's outer core density using the MSW effect, which is special for \dc\ setups. We have found that, depending on setup and $\stheta$, a precision on the outer core density as high as $0.5\%$ at $3\sigma$ can be reached for a superbeam and a neutrino factory. 

In conclusion, for small $\theta_{13}$, the beta beam may be the most promising beam option for the \dc\ detector, since it is a flavor-clean beam with moderate requirements for the detector. For large $\theta_{13}$, a superbeam or neutrino factory might be an interesting alternative, provided that an excellent energy resolution can be obtained and other detector requirements can be met. In particular, a low energy neutrino factory may benefit from the parametric enhancement and the core resonance energy of a few GeV. In addition, the intrinsic tau contamination can be avoided by the low beam energy. Finally, note that we have only studied parameter sets inspired by phase~I of such a proposal, whereas it is an option to increase the fiducial mass in a later phase. In this case, one should further optimize for fiducial mass in the range between 2~and 10~GeV instead of a lower threshold from the point of view of long-baseline physics. 

\subsubsection*{Acknowledgments}

We are thankful to Doug Cowen, Tyce DeYoung, Jason Koskinen, and Marek Kowalski for illuminating discussions, and to Teresa Montaruli for useful suggestions. This work has been supported by Deutsche
Forschungsgemeinschaft, contracts WI 2639/2-1  [J.T, W.W.], WI 2639/3-1 and WI 2639/4-1 [W.W.], by the
DFG-funded research training group  1147 ``Theoretical astrophysics and
particle physics'' [J.T.], and by the European Union under the
European Commission Framework Programme~07 Design Study EUROnu,
Project 212372.

\begin{appendix}
\section{Energy-dependent effective fiducial volume}
\label{app:fidvol}

The effective area or fiducial volume of a neutrino telescope depends on the neutrino energy, since a larger energy produces a longer muon track or cascade, and thus deposits more light. In the extreme case of low energies, only the volume close to the photomultipliers will contribute to the fiducial volume, whereas for very high energies even light from outside of the actual detector volume may be detected. In the main text of this study, we parameterize the fiducial volume by a fixed value with a threshold. Here we show the effects of an energy-dependent fiducial volume generated for a preliminary simulation of \dc\ phase~I~\cite{Koskinen}. 
Note that while the conclusions in the main text do not depend on the actual geometry of the detector, the conclusions in this Appendix depend on the fiducial volumes.

\begin{figure}[!t]
 \centering
 \includegraphics[width=0.9\textwidth]{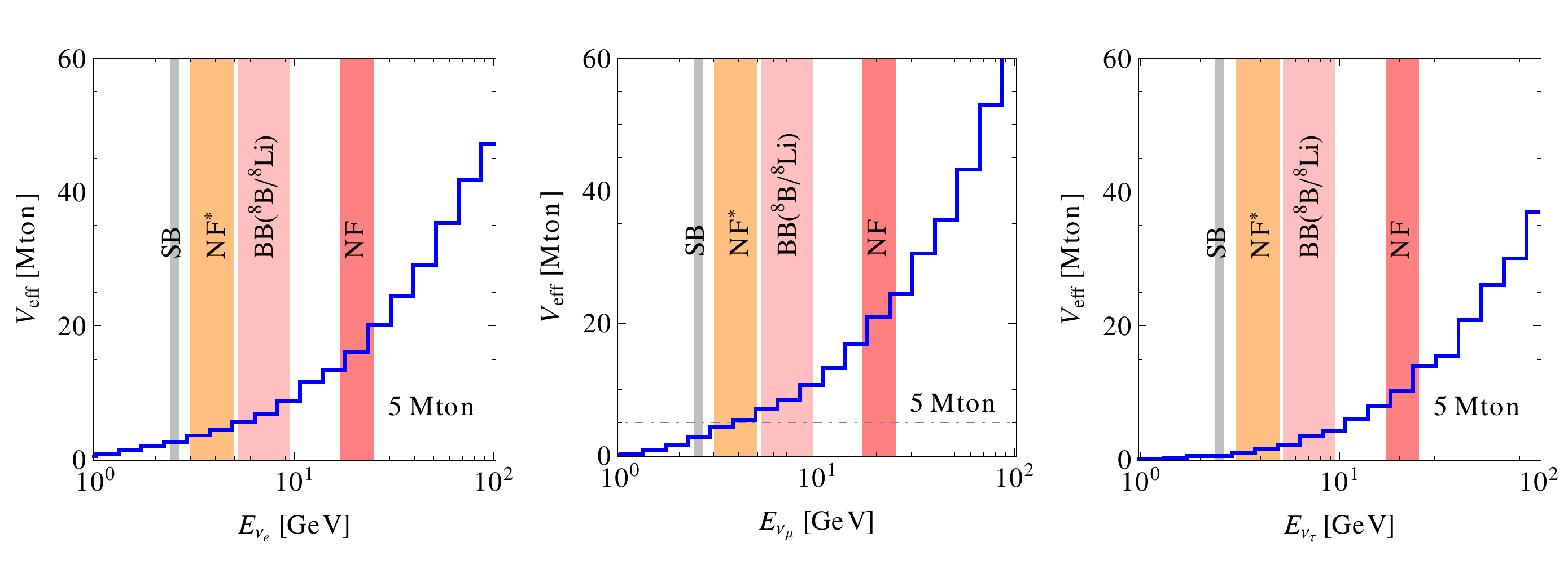}
 \mycaption{The effective fiducial volume of a preliminary \dc\ configuration as a function of the incoming neutrino energy for three flavor neutrinos. The dashed-dotted line is the constant fiducial volume chosen for the benchmark setups in the main text. The peak energies of the different neutrino sources are marked. 
}
 \label{fig:DC-Veff}
\end{figure}

We show the fiducial volume as a function of energy for the different flavors in \figu{DC-Veff}. This volume is defined as
\begin{equation}
 V_{\mathrm{eff}} = \frac{\text{Events triggered}}{\text{Events generated}} \times \text{Generation volume} \, ,
\label{equ:vfid}
\end{equation}
\ie, it has to be interpreted as a function of the reconstructed neutrino energy. Since it includes charged current and neutral current interactions, and it does not include cuts required to improve the flavor identification, and can only be regarded as an estimate.  We also show in \figu{DC-Veff} the peak energies of the different instruments ($E_\mu=25 \, \mathrm{GeV}$ for the neutrino factory), and the value for the fiducial volume used as standard value in the main text (5~Mt). One can easily extract from this figure (middle panel) that our estimate is a good match for the low energy neutrino factory denoted by NF$^*$ and beta beam denoted by BB($^8$B/$^8$Li). However, the fiducial volume for the high energy neutrino factory, denoted by NF, is actually significantly underestimated
and the effective volume for the superbeam (left panel) is slightly overestimated. Note that the beam has a high energy tail, which may be useful in parts of the parameter space. 
From the right panel ($\nu_\tau$), one can also extract that our $\nu_\tau$ contaminations, which are based on the same fiducial volume as $\nu_e$ and $\nu_\mu$ in the main text, are somewhat overestimated.

\begin{figure}[!t]
 \centering
 \includegraphics[width=0.9\textwidth]{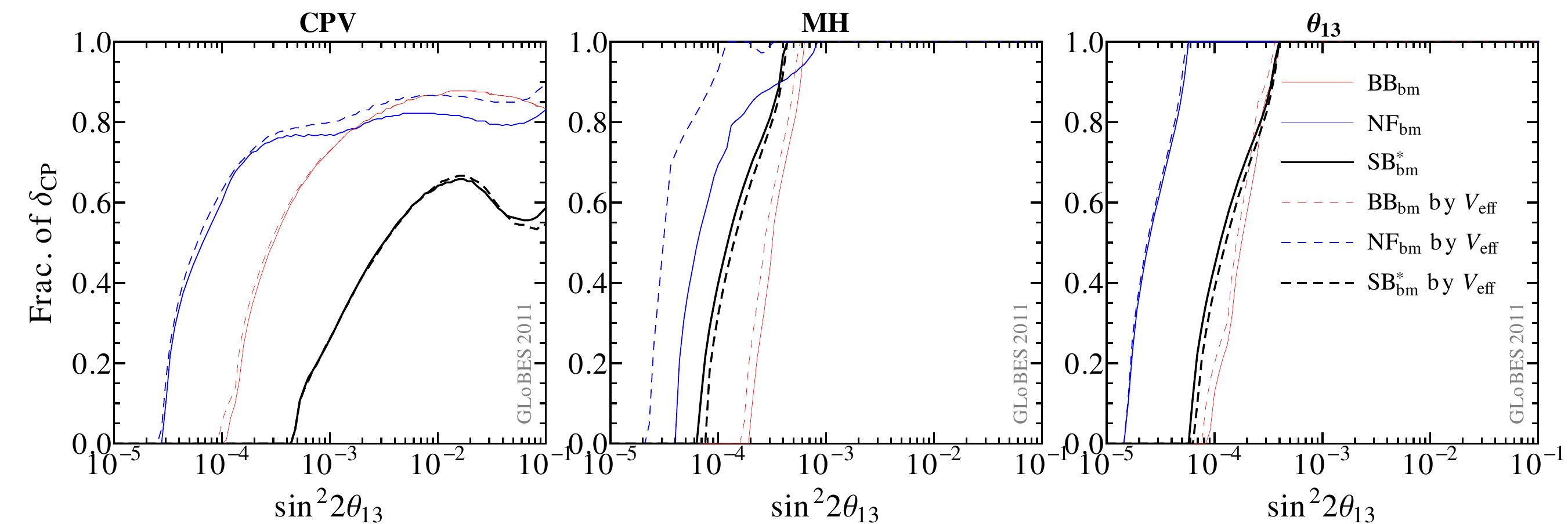}
 \mycaption{A comparison of the discovery reach of CPV, MH and $\theta_{13}$ at $3\sigma$ for the benchmark setups (\cf, \Tab~\ref{tab:standard}) with or without the effective volume $V_{\mathrm{eff}}$ of \figu{DC-Veff}. The benchmark setups without energy-dependent $V_{\mathrm{eff}}$ are represented by the solid curves, while those with energy-dependent $V_{\mathrm{eff}}$ are represented by the dashed curves. Here we assume the normal hierarchy.}
 \label{fig:Veff}
\end{figure}

Finally, in \figu{Veff}, a quantitative comparison between the setups based on fiducial volume and threshold in \Tab~\ref{tab:standard} and those based on \figu{DC-Veff} are shown in (two-baseline setups for beta beam and neutrino factory used). This figure confirms the previous estimates for the full simulation.

\end{appendix}

\bibliographystyle{apsrev}
\bibliography{references}

\end{document}